\documentclass[pra,preprint,superscriptaddress]{revtex4-1}

\def\ep{\epsilon_}
\def\e{{\rm e}}
\def\g{\boldsymbol}

\def\scr{\scriptstyle}

\usepackage{graphics}
\usepackage[dvips]{graphicx}	
\usepackage[english]{babel}
\usepackage{booktabs}
\usepackage{amsmath}
\usepackage{amssymb}
\usepackage{amscd}
\usepackage{array}
\usepackage{supertabular}
\usepackage{subfigure}
\DeclareTextSymbol{\degre}{OT1}{23}

\numberwithin{equation}{section}

\begin{document}
\title{Nonperturbative treatment of coupled anharmonic vibrational molecular modes analyzed by 2D-infrared spectroscopy}
\author{Kuo Kan Liang}\email{kkliang@sinica.edu.tw}
\affiliation{Research Center for Applied Sciences, Academia Sinica, Taipei 115, Taiwan}
\affiliation{Department of Biochemical Science and Technology, National Taiwan University, Taipei 106, Taiwan}
\author{Albert A. Villaeys}
\affiliation{Research Center for Applied Sciences, Academia Sinica, Taipei 115, Taiwan}
\affiliation{Universit\'e de Strasbourg and Institut de Physique et Chimie des Mat\'eriaux de Strasbourg, France}
\begin{abstract}
In the present work, we give an analytical non-perturbative treatment of mode-mode coupling and anharmonicity occurring in molecular vibrational systems analyzed by 2D-infrared spectroscopy. This analytical description allows a detailed discussion of the intricate structure of the 2D-infrared spectra resulting from the contributions provided by the various possible chronological orderings of the interactions with fields due to overlaps of the pulses. The peculiar role of the vibrational overtones and other combination states on the resonance shapes is elucidated and conditions for a better quantitative analysis are discussed.   
\end{abstract}
\maketitle 

\section{Introduction}

Taking advantage of the enormous strides made in recent years in the better control of infrared laser fields, two-dimensional (2D) vibrational infrared spectroscopy  has become a major tool in probing the dynamics induced by optical excitations and taking place among the vibrational coordinates driven by their corresponding force constants \cite{bb01,bb02,bb03}. 
Traditional vibrational spectroscopy has failed in disentangling unambiguously the information provided either by the various fundamental, overtone and combination band transition frequencies on the ground state nuclear potential surface, or by the vibrational line shapes like energy, phase or orientational relaxation. Analogously, linear optical spectroscopy cannot differentiate unambiguously between the limiting cases of fast and slow modulations, while various nonlinear optical spectroscopies such as fluorescence line narrowing, hole burning, or photon-echo are quite sensitive to the nature of the line broadening \cite{bb04}.
A better understanding of the molecular interactions, and their subsequent internal dynamics in condensed phase, is attainable today using 2D-infrared spectroscopy which makes it possible to simultaneously follow the dynamics in different degrees of freedom \cite{bb05,bb06,bb07,bb08,bb09,bb10,bb11,bb12,bb13}.
This type of spectroscopy implies a high time resolution of the third-order coherent signal emitted by the molecular systems as a function of both detection time and pulse delay times, although a frequency analysis is done at the end. This is achieved by a double Fourier transform with respect to a pair of time variables among detection time and pulse delay times and leads to a 2D-frequency mapping whose resonance structure contains all the informations about dynamics and structure of the molecular systems. An efficient method to evaluate this 2D-frequency mapping has been developed more recently \cite{bb14,bb15,bb16,bb17}. It has been used to analyze the role of vibrations coupled to electronic motion and clarifies the processes underlying the existence of cross-peaks in 2D-spectra, like inter-mode couplings, anharmonicity and structure parameters.

Of course, other multi-dimensional nonlinear vibrational spectroscopies like infrared photon-echo \cite{bb18,bb19,bb20,bb21}, Raman echo \cite{bb22,bb23},
or transient IR pump-probe experiments \cite{bb24,bb25,bb26}, have been extensively developed in the past and are effective in disentangling various contributions to the vibrational line shape. They give insight into vibrational couplings. 
As an example, multidimensional spectroscopy based on off-resonant excitation by a series of femtosecond laser pulses, provides a direct and distinguishable information with respect to the homogeneous or inhomogeneous spectral line broadening. Notice that in general, inhomogeneous broadening implying statistical distributions do not carries dynamical information. Among the large number of nonlinear spectroscopies \cite{bb27,bb28},
2D-infrared spectroscopy is well adapted for measuring vibrational couplings, discriminating different broadening mechanisms or identifying transient structures on a picosecond time scale. Coherent 2D-infrared vibrational spectroscopy enables  not only to characterize vibrational couplings, but also to identify and quantify mechanisms of vibrational line width. Taking advantage of these nonlinear coherent spectroscopic methods, it is now possible to get physical insights on the potential energy surface in condensed phase with a time resolution of a few picoseconds, a time scale of particular interest in biological systems.

During the last decade, 2D-infrared spectroscopy has proved to be quite efficient for analyzing transient molecular structures \cite{bb29,bb30,bb31,bb32}.
More specifically, experiments suggested the peculiar roles of anharmonicity in the ground state, nonlinear dependence of the transition dipole moment on the normal coordinates, or likewise level-dependent dephasing dynamics. At least one of these processes is required to observe infrared optical nonlinearities \cite{bb05,bb33,bb34,bb35}. 
In fact, it has been shown that 2D-infrared vibrational spectroscopy has the ability of analyzing these different mechanisms from the entire characterization of the 2D-spectral resonances, involving their positions, amplitudes and shapes.

In most of the spectroscopy experiments we are dealing with, molecular systems contain many levels whose energies fit within the spectral ranges of the laser pulses.
In our description, because the pulses are described analytically, all the contributions from the successive transitions leading to the third-order polarization are treated on the same footing and properly weighted by their corresponding spectral overlaps of the pulses and resonances. Here, we describe a 2D-infrared spectroscopy experiment performed on a molecular system made of two oscillators mutually coupled anharmonically. Because of the strong inter mode coupling, a complete re-diagonalization is required. It implies not only a complete mixing of the energy levels but also a complete redistribution of the dipole moments as well as of the relaxation, transition and dephasing constants among the vibrational eigenstates. However, orientational effects are not considered here. Since orientational motions of molecules in condensed phase are usually slower than vibrational motions, the orientational dynamics can be factorized from the vibrational dynamics and described independently. Notice that an extensive study of the role played by the orientational motion in third-order optical responses has been considered by Golonzka {\it et al} \cite{bb36}.

In Sec.2, we give a formal description of the four-wave mixing (4WM) signal obtained by heterodyne detection in the phase-matched direction corresponding to the usual photon-echo (PE) detection direction. At the end, a double Fourier transform is performed analytically over the detection time and one of the pulse delay times to reveal the resonance structure of the 2D-spectra encompassing all the information about the vibrational structure. Next, in Sec.3, we introduce the model of coupled anharmonic oscillators describing the vibrational system under investigation. In the present model, all internal couplings are exactly accounted for by a proper diagonalization of the levels, dipole moments and various relaxation constants. Also, the evolution Liouvillian of the coherences and populations are evaluated, for they participate in the third-order polarization. Finally, in Sec.4, some simulations are performed to emphasize the influence of anharmonicity and inter-mode couplings on the 2D-infrared spectrum. Also, the peculiar role played by the various non-rephasing contributions associated with the vibrational overtones and other combination states are investigated. For convenience, in the numerical simulation, the frequency range has been limited to the first two transitions but the dynamics of the higher states are, of course, retained in the calculation.

\section{Formal description of the 4WM photon-echo signal underlying its subsequent 2D-infrared Fourier spectrum} 

Most of the 2D-infrared spectra, obtained from the recently developed 2D-spectroscopy, are based on the double Fourier transformation of the usual 4WM photon-echo signal 
obtained by the interaction of three exciting laser fields with wave vectors $\vec{k}_a$, $\vec{k}_b$ and $\vec{k}_c$ acting on a molecular system. The resulting signal, termed 4WM photon-echo, is the four-wave mixing signal obtained by heterodyne detection in the phase-matched direction $-\vec{k}_a+\vec{k}_b+\vec{k}_c$ corresponding to the photon-echo geometry. The vibrational molecular system of interest here is made of two vibrational oscillators undergoing anharmonicity and mode-mode coupling, and will be described in the next section.

The interaction Hamiltonian between the vibrational molecular system and the laser beams is given by%
\begin{eqnarray}
\g{V}(t)=-\sum_{p=a,b,c} \mathcal{A}_p(t-T_p)\bigl[\vec{\g{\mu}}\cdot\vec{E}_p\e^{-i\omega_p(t-T_p)+i\vec{k}_p\cdot\vec{r}}+ {\rm C.C.}\bigr],
\label{2.1}
\end{eqnarray}    
where the notation C.C. stands for the complex conjugate part and $\mathcal{A}_p(t-T_p)$ represents the normalized envelop of the pulsed laser field $p$ given by
\begin{eqnarray}
\mathcal{A}_p(t-T_p)=\sqrt{\gamma_p}\exp(-\gamma_p\vert t-T_p\vert),  
\label{2.2}
\end{eqnarray}   
where $T_p$ and $\gamma_p^{-1}$ are the probing time and the duration of the pulse, respectively. Here, the laser pulse is described by a double-sided exponential, because this is the only analytical shape which allows a complete analytical evaluation of the polarization. As usual, $\vec{\g{\mu}}$ and $\omega_p$ are standard notations for the dipole moment operator of the vibrational system and frequency of the laser field $p$, respectively. It is well established that 4WM photon-echo and many other related nonlinear optical processes \cite{bb17,bb27,bb28,bb29,bb30,bb37,bb38,bb39,bb40,bb41,bb42,bb43,bb44,bb45,bb46,bb47,bb48,bb49} are accounted for by the third-order perturbation term of the density matrix with respect to the laser-molecule interaction $\g{V}(t)$ described previously. For the three-pulse process we want to discuss here, the contribution to the third-order term of the density matrix $\g{\rho}^{(3)}(t)$ takes the form
\begin{multline}
\g{\rho}^{(3)}(t)=\frac{i}{\hbar^3}\int_{t_0}^{t}\!\!\!d\tau_3\int_{t_0}^{\tau_3}\!\!\!d\tau_2\int_{t_0}^{\tau_2}\!\!\!d\tau_1\g{G}(t-\tau_3)\g{L}_{\rm v}(\tau_3)\g{G}(\tau_3-\tau_2)\g{L}_{\rm v}(\tau_2)\g{G}(\tau_2-\tau_1) \g{L}_{\rm v}(\tau_1)\g{\rho}(t_0).  
\label{2.3}
\end{multline}   
The various $\g{G}(\tau_i-\tau_j)$ are the free evolution Liouvillians of the vibrational system alone, corresponding to $\g{G}(\tau_i-\tau_j)=\e^{-\frac{i}{\hbar}\g{L}(\tau_i-\tau_j)}$ where $\g{L}=[\g{H},\cdots]$ with $\g{H}$ the Hamiltonian of the vibrational molecular system. The matrix elements of the type $\g{G}_{mmmm}(\tau_i-\tau_j)$ accounts for the population evolution of the state $m$, while the ones of the type $\g{G}_{mnmn}(\tau_i-\tau_j)$ with $m\not= n$ accounts for the evolution of the coherence between states $m$ and $n$. Notice that all free evolution Liouvillian matrix elements of the type $\g{G}_{mnmn}(\tau_i-\tau_j)$ are diagonal in the coherence Liouvillian subspace and they have the general form
\begin{eqnarray}
 \g{G}_{mnmn}(\tau_i-\tau_j)=\e^{-i\omega_{mn}(\tau_i-\tau_j) -\g{\Gamma}_{mnmn}(\tau_i-\tau_j)}, 
\label{2.4}
\end{eqnarray}
while the ones of the type $\g{G}_{mmmm}(\tau_i-\tau_j)$ need to be evaluated on a specific model. Also, the definition of the interaction Liouvillian is given by $\g{L}_{{\rm v},\,ijkl}(\tau)=\g{V}_{ik}(\tau)\delta_{lj}-\g{V}_{lj}(\tau)\delta_{ik}$. The density matrix elements required for the evaluation of the third-order polarization are obtained from the set of pathways contributing to the third-order process. They will be evaluated from a vibrational model appropriate for our purpose and introduced in the next section.

The formal expression of the third-order polarization in the phase-matched direction $-\vec{k_a}+\vec{k_b}+\vec{k_c}$, required in the model of 4WM photon-echo experiment, is given in the eigenstate basis set $\{\vert\ep j\rangle\}$ of the vibrational system by
\begin{eqnarray}
\vec{\g{P}}_{\scr{-\vec{k_a}+\vec{k_b}+\vec{k_c}}}^{(3)}(\vec{r},t)=\sum_{\ep i,\ep j}\g{\rho}_{\ep i\ep j}^{(3)}(t)\vec{\g{\mu}}_{\ep j\ep i},
\label{2.5}
\end{eqnarray} 
where, from Eqs.(\ref{2.2}) and (\ref{2.3}), the density matrix elements can be written as
\begin{multline}
\g{\rho}_{\ep i\ep j}^{(3)}(t)=\frac{i}{\hbar^3}\sum_{\{n\}}^{'}\sum_{r,q,p}\int_{t_0}^{t}d\tau_3\int_{t_0}^{\tau_3}d\tau_2\int_{t_0}^{\tau_2}d\tau_1 \\
\times\mathcal{A}_r(\tau_1-T_r)\mathcal{A}_q(\tau_2-T_q)\mathcal{A}_p(\tau_3-T_p)\g{R}_{n,\ep i\ep j}(\tau_1,\tau_2,\tau_3,t)
\e^{-i(\vec{k}_a-\vec{k}_b-\vec{k}_c)\cdot\vec{r}}.
\label{2.6}
\end{multline}  
Each $n$ specifies a particular pathway in the Liouvillian space contributing to the density matrix elements.
The symbol $\sum_{r,q,p}$  stands for the summation over the field combinations. These combinations must satisfy the phase-matching condition in the framework of the rotating wave approximation where just the combinations of field components satisfying the secular approximation are retained. In addition, the symbol $\sum_{\{n\}}^{'}$ means that only the values of $n$ associated with the particular density matrix element $\g{\rho}_{\ep i\ep j}^{(3)}(t)$ must be retained. Finally, the general mathematical structure of $\g{R}_{n,\ep i\ep j} (\tau_1, \tau_2,\tau_3,t)$ is of the type
\begin{eqnarray}
\g{R}_{n,\ep i\ep j}(\tau_1,\tau_2,\tau_3,t)= Q_{n,r,q,p}\e^{A_{n,r,q,p}\tau_3+B_{n,r,q}\tau_2+C_{n,r}\tau_1}\e^{K_{n,r,q,p}t},
\label{2.7}
\end{eqnarray} 
where all the amplitudes $Q_{n,r,q,p}$ and exponential arguments $A_{n,r,q,p}$, $B_{n,r,q}$, $C_{n,r}$, and $K_{n,r,q,p}$ will be obtained from identifying \eqref{2.7} with the matrix elements $\langle \epsilon_i\vert\g{G}(t-\tau_3)\g{L}_{\rm v}(\tau_3)\g{G}(\tau_3-\tau_2)\g{L}_{\rm v}(\tau_2)\g{G}(\tau_2-\tau_1) \g{L}_{\rm v}(\tau_1)\g{\rho}(t_0)\vert\epsilon_j\rangle$ associated with pathway $n$. They will be evaluated in the following from our specific model. 

To preserve the phase information contained in the third-order polarization, which is built from successive excitation by the laser pulses, we use heterodyne detection by mixing the third-order signal with a local field oscillator $\vec{E}_{\rm lo}(\vec{r},\tau)$ which has the form of
\begin{eqnarray}
\vec{E}_{\rm lo}(\vec{r},\tau)=\mathcal{A}_{\rm lo}(\tau-t)\bigl[\vec{E}_{\rm lo}(\omega_{\rm lo})\e^{-i\omega_{\rm lo}\tau+i\vec{k}_{\rm lo}\cdot\vec{r}+i\Psi}+ {\rm C.C.}\bigr]
\label{2.8}
\end{eqnarray} 
with adjustable phase $\Psi$, and spatial and temporal overlaps. For simplicity, the envelop of the local oscillator pulse is chosen much shorter than all the other characteristic times. Also, the local field amplitude is chosen much larger than that of the 4WM photon-echo signal. Then, by subtracting the intensity of the local field oscillator and rejecting the high frequency terms, one finally obtains the detected 4WM photon-echo signal
\begin{eqnarray}
I_{\rm 4WM-pe}(t)\approx 4 \Re\Bigl\lbrace i \vec{E}_{\rm lo}^{\star}(\omega_{\rm lo})\cdot\vec{\g{P}}_{\scr{-\vec{k_a}+\vec{k_b}+\vec{k_c}}}^{(3)}(\vec{r},t)\e^{i\omega_{\rm lo}t-i\vec{k}_{\rm lo}\cdot\vec{r}-i\Psi}\Bigr\rbrace.
\label{2.9}
\end{eqnarray} 
In addition, with the convenient choices of the wave vector $\vec{k}_{\rm lo}=-\vec{k_a}+\vec{k_b}+\vec{k_c}$ and frequency $\omega_{\rm lo}=-\omega_a+\omega_b+\omega_c$, appropriate for the phase-matching conditions chosen in this experiment, one can get 
\begin{eqnarray}
I_{\rm 4WM-pe}(t)\approx -4 \Im\Bigl\lbrace \vec{E}_{\rm lo}^{\star}(\omega_{\rm lo})\cdot\vec{\g{P}}_{\scr{-\vec{k_a}+\vec{k_b}+\vec{k_c}}}^{(3)}(\vec{r},t)\e^{i\omega_{\rm lo}t-i\vec{k}_{\rm lo}\cdot\vec{r}-i\Psi}\Bigr\rbrace 
\label{2.10}
\end{eqnarray} 
which in turn, on account of the expression (\ref{2.6}), gives the result
\begin{multline}
I_{\rm 4WM-pe}(t)\approx -\frac{4}{\hbar^3}\Re \Bigl\{\vec{E}_{\rm lo}^{\star}(\omega_{\rm lo})\cdot\vec{\g{\mu}}_{\ep j\ep i}
\,\e^{i\omega_{\rm lo}t-i\Psi}\sum_{\ep i\ep j}\sum_{\underset{\vec{k}_r+\vec{k}_q +\vec{k}_p=-\vec{k}_a+\vec{k}_b+\vec{k}_c}{r,q,p}} \\
\times\int_{t_0}^{t}d\tau_3\int_{t_0}^{\tau_3}d\tau_2\int_{t_0}^{\tau_2}d\tau_1\mathcal{A}_r(\tau_1-T_r)\mathcal{A}_q(\tau_2-T_q)
\mathcal{A}_p(\tau_3-T_p)\g{R}_{n,\ep i\ep j}(\tau_1,\tau_2,\tau_3,t)\Bigr\}.
\label{2.11}
\end{multline} 
It cannot be overemphasized that the advantage of using an heterodyne detection lies in the possibility of testing the real and the imaginary parts of the polarization by a convenient choice of the local oscillator phase. Assuming the dipole moments as well as the amplitude of the local field oscillator $\vec{E}_{\rm lo}^{\star}(\omega_{\rm lo})$ to be real, we have
\begin{multline}
I_{\rm 4WM-pe}(t)=
-\frac{4}{\hbar^3}\sum_{\ep i\ep j}\vec{E}_{\rm lo}(\omega_{\rm lo})\cdot\vec{\g{\mu}}_{\ep j\ep i}\sum_{\underset{\vec{k}_r+\vec{k}_q +\vec{k}_p=-\vec{k}_a+\vec{k}_b+\vec{k}_c}{r,q,p}}\int_{t_0}^{t}d\tau_3\int_{t_0}^{\tau_3}d\tau_2\int_{t_0}^{\tau_2} d\tau_1                            \\
\times\Biggl\{\Re\Bigl[\e^{i\omega_{\rm lo}t}\mathcal{A}_r(\tau_1-T_r)\mathcal{A}_q(\tau_2-T_q)
\mathcal{A}_p(\tau_3-T_p)\g{R}_{n,\ep i\ep j}(\tau_1,\tau_2,\tau_3,t)\Bigr]\delta_{\Psi,0}                                                              \\
+\Im\Bigl[\e^{i\omega_{\rm lo}t}\mathcal{A}_r(\tau_1-T_r)\mathcal{A}_q(\tau_2-T_q)
\mathcal{A}_p(\tau_3-T_p) \g{R}_{n,\ep i\ep j}(\tau_1,\tau_2,\tau_3,t)\Bigr]\delta_{\Psi,\frac{\pi}{2}}\Biggr\},
\label{2.12}
\end{multline} 

Finally, to get the explicit expression of the photon-echo signal, the successive time-integrations must be performed analytically. The results are presented below. The first time-integration gives the general expression
\begin{eqnarray}
\int_{-\infty}^{t_j}dt_i\mathcal{A}_r(t_i-T_r)\e^{Xt_i}=H(T_r-t_j)P_i[1]\e^{p_i[1]t_j}+H(t_j-T_r)\sum_{\alpha=1}^{2}P_s[\alpha]\e^{p_s[\alpha]t_j}
\label{2.13} 
\end{eqnarray}
where, for convenience, the various constants are presented in Appendix A. Next, the second time-integration can be expressed as
\begin{align}
\int_{-\infty}^{t_k}&dt_j\mathcal{A}_q(t_j-T_q)\e^{Yt_j}\int_{-\infty}^{t_j}dt_i\mathcal{A}_r(t_i-T_r)\e^{Xt_i}\nonumber\\
=&H(T_r-T_q)\bigl[H(T_q-t_k)D_{i-rsq}[1]\e^{d_{i-rsq}[1]t_k}
+H(t_k-T_q)H(T_r-t_k)D_{m-rsq}[1]\e^{d_{m-rsq}[1]t_k} \nonumber\\
&+\sum_{\alpha=1}^{4}H(t_k-\tau_{s-rsq}[\alpha])D_{s-rsq}[\alpha]\e^{d_{s-rsq}[\alpha]t_k}\bigr]
+H(T_q-T_r)\bigl[H(T_r-t_k)D_{i-qsr}[1]\e^{d_{i-qsr}[1]t_k}\nonumber\\
&+H(t_k-T_r)H(T_q-t_k)\sum_{\alpha=1}^{2}D_{m-qsr}[\alpha]\e^{d_{m-qsr}[\alpha]t_k}\nonumber\\
&+\sum_{\alpha=1}^{4}H(t_k-\tau_{s-qsr}[\alpha])D_{s-qsr}[\alpha]\e^{d_{s-qsr}[\alpha]t_k}\bigr]
\label{2.14} 
\end{align}
In this expression, the two different chronological orderings $T_r>T_q$ and $T_q>T_r$ of the laser field interactions are both accounted for. The corresponding constants are also described in Appendix A. Finally, the third time-integration can be presented as
{\small
\begin{align}
\int_{-\infty}^{t}dt_k&\mathcal{A}_p(t_k-T_p)\e^{Zt_k}\int_{-\infty}^{t_k}dt_j\mathcal{A}_q(t_j-T_q)\e^{Yt_j}\int_{-\infty}^{t_j}dt_i
\mathcal{A}_r(t_i-T_i)\e^{Xt_i}\nonumber\\
=&H(T_p-T_q)H(T_q-T_r)\Bigg\{H(\tau_{i-pqr}-t)F_{i-pqr}[1]\e^{f_{i-pqr}[1]t}+\sum_{\alpha=1}^6 H(t-\tau_{m-pqr-1}[\alpha])\nonumber\\
&\times H(\tau_{m-pqr-2}[\alpha]-t)F_{m-pqr}[\alpha]\e^{f_{m-pqr}[\alpha]t}+\sum_{\alpha=1}^7 H(t-\tau_{s-pqr}[\alpha])F_{s-pqr}[\alpha]\e^{f_{s-pqr}[\alpha]t}\Bigg\}\nonumber\\
&+H(T_q-T_p)H(T_p-T_r)\Bigg\{H(\tau_{i-qpr}-t)F_{i-qpr}[1]\e^{f_{i-qpr}[1]t}+\sum_{\alpha=1}^4 H(t-\tau_{m-qpr-1}[\alpha])\nonumber\\
&\times H(\tau_{m-qpr-2}[\alpha]-t) F_{m-qpr}[\alpha]\e^{f_{m-qpr}[\alpha]t}+\sum_{\alpha=1}^8 
H(t-\tau_{s-qpr}[\alpha])F_{s-qpr}[\alpha]\e^{f_{s-qpr}[\alpha]t}\Bigg\}\nonumber\\
&+H(T_p-T_r)H(T_r-T_q)\Bigg\{H(\tau_{i-prq}-t)F_{i-prq}[1]\e^{f_{i-prq}[1]t}+\sum_{\alpha=1}^5 H(t-\tau_{m-prq-1}[\alpha])\nonumber\\
&\times H(\tau_{m-prq-2}[\alpha]-t) F_{m-prq}[\alpha]\e^{f_{m-prq}[\alpha]t}+\sum_{\alpha=1}^7 H(t-\tau_{s-prq}[\alpha])F_{s-prq}[\alpha]\e^{f_{s-prq}[\alpha]t}\Bigg\}\nonumber\\
&+H(T_r-T_p)H(T_p-T_q)\Bigg\{H(\tau_{i-rpq}-t)F_{i-rpq}[1]\e^{f_{i-rpq}[1]t}+\sum_{\alpha=1}^3 H(t-\tau_{m-rpq-1}[\alpha])\nonumber\\
&\times H(\tau_{m-rpq-2}[\alpha]-t) F_{m-rpq}[\alpha]\e^{f_{m-rpq}[\alpha]t}+\sum_{\alpha=1}^7 H(t-\tau_{s-rpq}[\alpha])F_{s-rpq}[\alpha]\e^{f_{s-rpq}[\alpha]t}\Bigg\}\nonumber\\
&+H(T_q-T_r)H(T_r-T_p)\Bigg\{H(\tau_{i-qrp}-t)F_{i-qrp}[1]\e^{f_{i-qrp}[1]t}+\sum_{\alpha=1}^3 H(t-\tau_{m-qrp-1}[\alpha])\nonumber\\
&\times H(\tau_{m-qrp-2}[\alpha]-t) F_{m-qrp}[\alpha]\e^{f_{m-qrp}[\alpha]t}+\sum_{\alpha=1}^7 H(t-\tau_{s-qrp}[\alpha])F_{s-qrp}[\alpha]\e^{f_{s-qrp}[\alpha]t}\Bigg\}\nonumber\\
&+H(T_r-T_q)H(T_q-T_p)\Bigg\{H(\tau_{i-rqp}-t)F_{i-rqp}[1]\e^{f_{i-rqp}[1]t}+\sum_{\alpha=1}^2 H(t-\tau_{m-rqp-1}[\alpha])\nonumber\\
&\times H(\tau_{m-rqp-2}[\alpha]-t) F_{m-rqp}[\alpha]\e^{f_{m-rqp}[\alpha]t}+\sum_{\alpha=1}^7 H(t-\tau_{s-rqp}[\alpha])F_{s-rqp}[\alpha]\e^{f_{s-rqp}[\alpha]t}\Bigg\}\nonumber\\
\label{2.15} 
\end{align}}
Here, six different chronological orderings are involved in the final result and their corresponding contributions are labeled by the subscripts $pqr$ if $T_p>T_q>T_r$, $qpr$ if $T_q>T_p>T_r$, $prq$ if $T_p>T_r>T_q$, $rpq$ if $T_r>T_p>T_q$, $qrp$ if $T_q>T_r>T_p$ and $rqp$ if $T_r>T_q>T_p$, successively. For the sake of convenience, all the constants are listed in Appendix A.

Finally, when all the constants $X$, $Y$ and $Z$ associated with the various pathways required for the particular vibrational molecular system are known, the 2D-spectrum can be obtained by performing the double Fourier transformations over pulse delay time, here chosen as $T_a-T_b$, and the detection time $t$. It takes the general form
\begin{eqnarray}
I_{\rm 4WM-pe}(\omega_{\tau},\omega_{t})=\int_{-\infty}^{\infty}\,dt\e^{i\omega_{t}t} \int_{-\infty}^{\infty}\,d(T_a-T_b)\e^{i\omega_{\tau}(T_a-T_b)} I_{\rm 4WM-pe}(T_a-T_b,t) ,
\label{2.16} 
\end{eqnarray}
where we have introduced the explicit dependence of the 4WM photon-echo signal on the pulse delay time. From the calculation of the Fourier transforms, Eq.(\ref{2.12}) gives 
\begin{multline}
I_{\rm 4WM-pe}(t)= -\frac{4}{\hbar^3}\sum_{\ep i\ep j}\vec{E}_{\rm lo}(\omega_{\rm lo})\cdot\vec{\g{\mu}}_{\ep j\ep i}      
\Biggl\{\Re\Bigl[I_{pqa}(\omega_{\tau},\omega_{t})+I_{par}(\omega_{\tau},\omega_{t})+I_{aqr}(\omega_{\tau},\omega_{t})\Bigr]\delta_{\Psi,0}     \\
+\Im\Bigl[I_{pqa}(\omega_{\tau},\omega_{t})+I_{par}(\omega_{\tau},\omega_{t})+I_{aqr}(\omega_{\tau},\omega_{t})\Bigr]\delta_{\Psi,\frac{\pi}{2}}\Biggr\}.
\label{2.17}
\end{multline} 
An explicit evaluation will require the spectral decomposition obtained from the diagonalization of the vibrational Hamiltonian of the molecular system. It will be introduced in the next section. Besides, since in this experiment only the pulse delay times are relevant, one of the pulse probing time can be fixed arbitrarily as $T_b=0$.
The first contribution in the bracket, $I_{pqa}(\omega_{\tau},t)$, corresponds to
\begin{align}
I_{pqa}&(\omega_{\tau},\omega_{t})
=\frac{2\gamma_a^{\frac{3}{2}}\e^{i\omega_bT_b+i\omega_cT_c}}{(\omega_{\tau}-\omega_a)^2+\gamma_a^2} \sum_{q=\{b,c\}}\sum_{p=\{b,c\}}^{p\not=q}Q_{n,a,q,p}      \nonumber\\
\times&\Biggl[H(T_q-T_p)\Bigl[\frac{D^{pqa}_{i-qsp}[1]}{i\omega_{t}+d^{pqa}_{i-qsp}[1]+K_{n,a,q,p}}\e^{(i\omega_{t}+d^{pqa}_{i-qsp}[1]+K_{n,a,q,p})T_p}                    \nonumber\\
&+\frac{D^{pqa}_{m-qsp}[1]}{i\omega_{t}+d^{pqa}_{m-qsp}[1]+K_{n,a,q,p}}\bigl[\e^{(i\omega_{t}+d^{pqa}_{m-qsp}[1]+K_{n,a,q,p})T_q}-\e^{(i\omega_{t}+d^{pqa}_{m-qsp}[1] +K_{n,a,q,p})T_p}\bigr]                                                                                                                       \nonumber\\
&-\sum_{\beta=1}^{4}\frac{D^{pqa}_{s-qsp}[\beta]}{i\omega_{t}+d^{pqa}_{s-qsp}[\beta]+K_{n,a,q,p}}\e^{(i\omega_{t}+d^{pqa}_{s-qsp}[\beta]+K_{n,a,q,p})\tau^{pqa}_{s-qsp}[\beta]}\Bigr]                                                                                                                                             \nonumber\\
&+H(T_p-T_q)\Bigl[\frac{D^{pqa}_{i-psq}[1]}{i\omega_{t}+d^{pqa}_{i-psq}[1]+K_{n,a,q,p}}\e^{(i\omega_{t}+d^{pqa}_{i-psq}[1]+K_{n,a,q,p})T_q}      \nonumber\\
&+\sum_{\beta=1}^{2}\frac{D^{pqa}_{m-psq}[\beta]}{i\omega_{t}+d^{pqa}_{m-psq}[\beta]+K_{n,a,q,p}}\bigl[\e^{(i\omega_{t}+d^{pqa}_{m-psq}[\beta]+K_{n,a,q,p})T_p}-\e^{(i\omega_{t}+d^{pqa}_{m-psq}[\beta]+K_{n,a,q,p})T_q}\bigr]     \nonumber\\
&-\sum_{\beta=1}^{4}\frac{D^{pqa}_{s-psq}[\beta]}{i\omega_{t}+d^{pqa}_{s-psq}[\beta]+K_{n,a,q,p}}\e^{(i\omega_{t}+d^{pqa}_{s-psq}[\beta]+K_{n,a,q,p})\tau^{pqa}_{s-psq}[\beta]}\Bigr]
\Biggr].
\label{2.18} 
\end{align}
The second one is given by
\begin{align}
I_{par}&(\omega_{\tau},\omega_{t})=\frac{2\gamma_a^{\frac{3}{2}}\e^{i\omega_bT_b+i\omega_cT_c}}{(\omega_{\tau}-\omega_a)^2+\gamma_a^2}\sum_{r=\{b,c\}}\sum_{p=\{b,c\}}^{p\not=r}Q_{n,r,a,p}                                                                                                                                                             \nonumber\\
\times&\Biggl[H(T_r-T_p)\Bigl[ \frac{D^{par}_{i-rsp}[1]}{i\omega_{t}+d^{par}_{i-rsp}[1]+K_{n,r,a,p}}\e^{(i\omega_{t}+d^{par}_{i-rsp}[1]+K_{n,r,a,p})T_p}       \nonumber\\
&+\frac{D^{par}_{m-rsp}[1]}{i\omega_{t}+d^{par}_{m-rsp}[1]+K_{n,r,a,p}}\bigl[\e^{(i\omega_{t}+d^{par}_{m-rsp}[1]+K_{n,r,a,p})T_r}-\e^{(i\omega_{t}+d^{par}_{m-rsp}[1]+K_{n,r,a,p})T_p}\bigr]                                                                                                                                                 \nonumber\\
&-\sum_{\beta=1}^{5}\frac{D^{par}_{s-rsp}[\beta]}{i\omega_{t}+d^{par}_{s-rsp}[\beta]+K_{n,r,a,p}}\e^{(i\omega_{t}+d^{par}_{s-rsp}[\beta]+K_{n,r,a,p})\tau^{par}_{s-rsp}[\beta]}\Bigr]                                                                                                                                                      \nonumber\\
&+H(T_p-T_r)\Bigl[ \frac{D^{par}_{i-psr}[1]}{i\omega_{t}+d^{par}_{i-psr}[1]+K_{n,r,a,p}}\e^{(i\omega_{t}+d^{par}_{i-psr}[1]+K_{n,r,a,p})T_r}                   \nonumber\\
&+\sum_{\beta=1}^{3}\frac{D^{par}_{m-psr}[\beta]}{i\omega_{t}+d^{par}_{m-psr}[\beta]+K_{n,r,a,p}}\bigl[\e^{(i\omega_{t}+d^{par}_{m-psr}[\beta]+K_{n,r,a,p})T_p} -\e^{(i\omega_{t}+d^{par}_{m-psr}[\beta]+K_{n,r,a,p})T_r}\bigr]                                                                                        \nonumber\\
&-\sum_{\beta=1}^{5}\frac{D^{par}_{s-psr}[\beta]}{i\omega_{t}+d^{par}_{s-psr}[\beta]+K_{n,r,a,p}}\e^{(i\omega_{t}+d^{par}_{s-psr}[\beta]+K_{n,r,a,p})\tau^{par}_{s-psr}}\Bigr]\Biggr]  ,
\label{2.19} 
\end{align}
and the last one, $I_{aqr}(\omega_c,t)$, takes the form
\begin{align}
I_{aqr}&(\omega_{\tau},\omega_{t})
=\frac{2\gamma_a^{\frac{3}{2}}\e^{i\omega_bT_b+i\omega_cT_c}}{(\omega_{\tau}-\omega_a)^2+\gamma_a^2} \sum_{r=\{b,c\}}\sum_{q=\{b,c\}}^{q\not=r}Q_{n,r,q,a}      \nonumber\\
\times&\Biggl[H(T_r-T_q)\Bigl[\frac{D^{aqr}_{i-rsq}[1]}{i\omega_{t}+d^{aqr}_{i-rsq}[1]+K_{n,r,q,a}}\e^{(i\omega_{t}+d^{aqr}_{i-rsq}[1]+K_{n,r,q,a})T_q}         \nonumber\\
&+\frac{D^{aqr}_{m-rsq}[1]}{i\omega_{t}+d^{aqr}_{m-rsq}[1]+K_{n,r,q,a}}\bigl[\e^{(i\omega_{t}+d^{aqr}_{m-rsq}[1]+K_{n,r,q,a})T_r}-\e^{(i\omega_{t}+d^{aqr}_{m-rsq}[1] +K_{n,r,q,a})T_q}\bigr]                                                                                                                       \nonumber\\
&-\sum_{\beta=1}^{6}\frac{D^{aqr}_{s-rsq}[\beta]}{i\omega_{t}+d^{aqr}_{s-rsq}[\beta]+K_{n,r,q,a}}\e^{(i\omega_{t}+d^{aqr}_{s-rsq}[\beta]+K_{n,r,q,a})\tau^{aqr}_{s-rsq}[\beta]}\Bigr]                                                                                                                                             \nonumber\\
&+H(T_q-T_r)\Bigl[\frac{D^{aqr}_{i-qsr}[1]}{i\omega_{t}+d^{aqr}_{i-qsr}[1]+K_{n,r,q,a}}\e^{(i\omega_{t}+d^{aqr}_{i-qsr}[1]+K_{n,r,q,a})T_r}      \nonumber\\
&+\sum_{\beta=1}^{2}\frac{D^{aqr}_{m-qsr}[\beta]}{i\omega_{t}+d^{aqr}_{m-qsr}[\beta]+K_{n,r,q,a}}\bigl[\e^{(i\omega_{t}+d^{aqr}_{m-qsr}[\beta]+K_{n,r,q,a})T_q}-\e^{(i\omega_{t}+d^{aqr}_{m-qsr}[\beta]+K_{n,r,q,a})T_r}\bigr]     \nonumber\\
&-\sum_{\beta=1}^{6}\frac{D^{aqr}_{s-qsr}[\beta]}{i\omega_{t}+d^{aqr}_{s-qsr}[\beta]+K_{n,r,q,a}}\e^{(i\omega_{t}+d^{aqr}_{s-qsr}[\beta]+K_{n,r,q,a})\tau^{aqr}_{s-qsr}[\beta]}\Bigr]\Biggr].
\label{2.20} 
\end{align}
Thus, we have established the general analytical expression of the 2D-spectrum, valid for any vibrational system whose dynamics can be described by the general terms $\sum_{j}Q_j\e^{E_j}t$, implying that non-Markovian effects are not relevant. These results can be applied to a large variety of vibrational systems to characterize their structure as well as the specific processes occurring during the course of the experiment.

\section{Description of the vibrational molecular model and influence of the mode-mode coupling}

\begin{figure}[!ht]
\begin{center}
\includegraphics{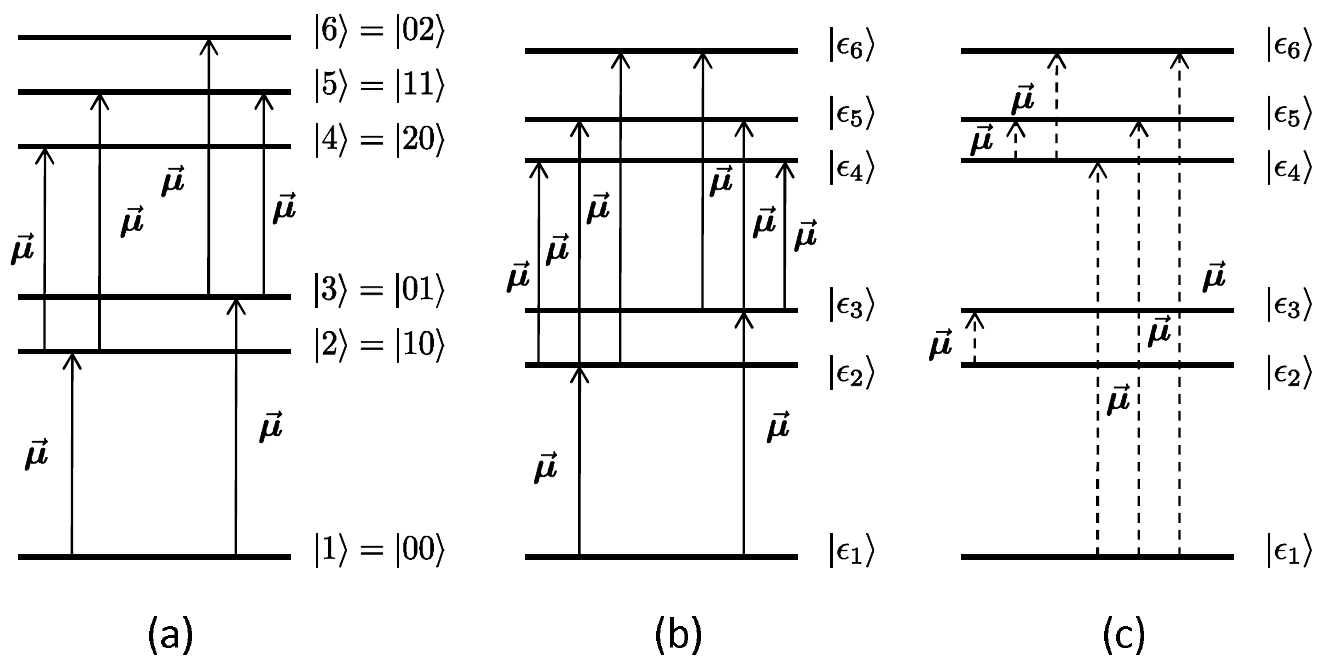} 
\end{center}
\caption{The states and dipole moments of the model vibrational system. In Panel (a), the two harmonic modes are independent and the states $\{\vert ij\rangle\}$ are obtained from the individual zeroth-order oscillator states. In Panel (b) and Panel (c), due to anharmonic effect and mode-mode couplings, the re-diagonalized eigenstates $\{\vert \epsilon_j\rangle\}$ are used for the representation of the anharmonic and coupled oscillators. The allowed dipole moments participating efficiently are indicated in (b), while the other allowed dipole moments shown in (c) have been neglected because of their negligible values.}
\label{fig_1}
\end{figure}

Most of the descriptions dedicated to the nonlinear optical responses of vibrational systems undergoing mode-mode couplings are handled perturbatively. Here, we are concerned with vibrational mode-mode couplings of a wide range of strength. For this purpose, the general framework adopted in our description must include an exact treatment of these couplings. It will enable, according to the strength of the couplings, a redistribution of the leading physical parameters driving the vibrational dynamics. This implies a redistribution of the dipole moments as well as all the dephasing, transition and decay constants.
For our purpose, the vibrational molecular model that is dealt with is made of two interacting harmonic oscillators undergoing anharmonicity and mode-mode coupling. Its corresponding Hamiltonian is made of three terms, 
\begin{eqnarray}
\g{H}=\g{H}_{{\rm v}_1}+\g{H}_{{\rm v}_2}+\g{U}(\g{Q}_1,\g{Q}_2),
\label{3.1}
\end{eqnarray} 
where $\g{H}_{{\rm v}_1}$ and $\g{H}_{{\rm v}_2}$ are the Hamiltonians of the non-interacting individual modes associated with the normal coordinates $\g{Q}_1$ and $\g{Q}_2$, while $\g{U}(\g{Q}_1,\g{Q}_2)$ describes their mutual interaction. As usual, among the various contributions associated with the normal mode expansion of the interaction Hamiltonian $\g{U}(\g{Q}_1,\g{Q}_2)$, the ones associated with one specific mode stand for anharmonicity  $\g{U}_{{\rm anh}}(\g{Q}_1,\g{Q}_2)$, while the other ones depending on both modes account for mode-mode coupling $\g{U}_{{\rm coupl}}(\g{Q}_1,\g{Q}_2)$. They are given by the expressions
\begin{eqnarray}
\g{U}(\g{Q}_1,\g{Q}_2)=\g{U}_{{\rm anh}}(\g{Q}_1,\g{Q}_2)+\g{U}_{{\rm coupl}}(\g{Q}_1,\g{Q}_2)
\label{3.2}
\end{eqnarray} 
where, from polynomial expansions on $\g{Q}_1$ and $\g{Q}_2$, we have
\begin{eqnarray}
&&\g{U}_{{\rm anh}}(\g{Q}_1,\g{Q}_2)=\frac{u_{111}}{6}\g{Q}_1^3+\frac{u_{222}}{6}\g{Q}_2^3                                          \nonumber\\
&&\g{U}_{{\rm coupl}}(\g{Q}_1,\g{Q}_2)=u_{12}\g{Q}_1\g{Q}_2+\frac{u_{122}}{2}\g{Q}_1\g{Q}_2^2+\frac{u_{112}}{2}\g{Q}_1^2\g{Q}_2.
\label{3.3}
\end{eqnarray} 
The matrix representation of the vibrational Hamiltonian in the basis of the individual harmonic modes involving ground state $\vert00\rangle$, singly excited states $\vert10\rangle,\vert01\rangle$, overtones $\vert20\rangle,\vert02\rangle$ and combination state $\vert11\rangle$, can be straightforwardly obtained from the expressions of the vibrational Hamiltonian including $\g{U}_{{\rm anh}}(\g{Q}_1,\g{Q}_2)$ and $\g{U}_{{\rm coupl}}(\g{Q}_1,\g{Q}_2)$. 
The perturbation expansion of the density matrix, required to get the nonlinear optical response of the vibrational system, is more conveniently expressed in the eigenstate basis set which is obtained from the simple diagonalization of the vibrational Hamiltonian given by
\begin{eqnarray}
\bigl[\g{H}_{{\rm v}_1}+\g{H}_{{\rm v}_2}+\g{U}_{{\rm anh}}(\g{Q}_1,\g{Q}_2)+\g{U}_{{\rm coupl}}(\g{Q}_1,\g{Q}_2)\bigr]\vert \epsilon_i\rangle=\epsilon_i\vert \epsilon_i\rangle,
\label{3.4}
\end{eqnarray} 
where $\epsilon_i$ and $\vert \epsilon_i\rangle$ stand for the eigenvalues and the eigenstates of the vibrational system.
If it is assume (without lost of generality) that $\omega_1<\omega_2$, from low to high energy, the order of the states is ${\vert 00\rangle,\vert 10\rangle,\vert 01\rangle,\vert 20\rangle,\vert 11\rangle,\vert 02\rangle}$. These states are the only ones accessible during the course of the photon-echo process of interest here. Also, the notation $\vert n m\rangle$ stands for $n$ and $m$ quanta in oscillators 1 and 2, respectively. 
All the vibrational states of the molecular system and their corresponding dipole moments are presented in Fig.~\ref{fig_1} for both the zero-order and the eigenstate representations.

After solving the eigenvalue problem, we have to derive the expressions of all the dephasing, transition, and damping constants, as well as the expressions of the matrix elements of the dipole moments in the diagonal basis set. These modified constants and matrix elements are the direct consequence of the redistribution induced by anharmonicity and mode-mode-coupling interactions. It is important to note that with the typical values of the physical parameters like the ones chosen for our numerical simulation, some of the matrix elements of the dipole moments are orders of magnitude smaller than the other ones. This is particularly the case for the transition dipole moments $\vec{\g{\mu}}_{\ep 2\ep 3}$,   $\vec{\g{\mu}}_{\ep 4\ep 5}$, $\vec{\g{\mu}}_{\ep 5\ep 6}$ and  $\vec{\g{\mu}}_{\ep 4\ep 6}$. Therefore, their contributions are not taken into account here, as shown in Fig.~\ref{fig_1}.
The transformation from the zero-order representation to the eigenstate representation can be obtained from the Liouvillian equation of the vibrational molecular system 
\begin{eqnarray}
\frac{\partial \g{\rho}(t)}{\partial t}=-\frac{i}{\hbar}\g{L}\g{\rho}(t)-\g{\Gamma}\g{\rho}(t),
\label{3.5}
\end{eqnarray}  
where $\g{L}$ is the Liouvillian with respect to the vibrational Hamiltonian $\g{H}=\g{H}_{{\rm v}_1}+\g{H}_{{\rm v}_2}+\g{U}(\g{Q}_1,\g{Q}_2)$. As usual, $\g{\Gamma}$ stands for the damping Liouvillian and $\g{\rho}(t)$ for the density operator of the two anharmonic and coupled vibrational modes. The transformation between the zeroth-order and eigenstate basis sets, is obtained by writing Eq.(\ref{2.5}) in both representations. By identifying corresponding terms, we get the matrix representation of the damping Liouvillian
\begin{eqnarray}
\g{\Gamma}_{\alpha\beta\nu\lambda}=\sum_{tupq}\langle\alpha\vert t\rangle\langle u\vert\beta\rangle\langle p\vert\nu\rangle\langle \lambda\vert q\rangle\g{\Gamma}_{tupq} .
\label{3.6}
\end{eqnarray}  
For the dipole moments, we have a similar relation given by
\begin{eqnarray}
\langle\epsilon_i\vert\vec{\g{\mu}}\vert\epsilon_j\rangle=\sum_{m,n}\langle\epsilon_i\vert m\rangle\langle n\vert \epsilon_j\rangle \vec{\g{\mu}}_{mn}.
\label{3.7}
\end{eqnarray}  
Notice that in previous expressions (\ref{3.6}) and (\ref{3.7}) the latin letters stand for the zero-order states while the greek letters correspond to the eigenstates. 
Once all these physical parameters driving the dynamics of the vibrational system are known, we can restate the dynamical evolution in the diagonal basis set $\{\vert \epsilon_j\rangle\}$ of the anharmonic and coupled vibrational modes.

The evaluation of the heterodyne-detected photon-echo signal and its subsequent 2D-infrared spectrum requires solving the dynamical equation of the vibrational molecular system in the eigenstate representation. The equations to solve can be expressed as
\begin{eqnarray}
\frac{\partial \g{\rho}_{\alpha\beta}(t)}{\partial t}&=& -\frac{i}{\hbar}\sum_{\nu\lambda}\g{L}_{\alpha\beta\nu\lambda}\g{\rho}_{\nu\lambda}(t)-\g{\Gamma}_{\alpha\beta\nu\lambda}\g{\rho}_{\nu\lambda}(t).
\label{3.8}
\end{eqnarray}  
To evaluate the evolutions of coherence and population, first we need the expression of the damping Liouvillian in the eigenstates representation and it can be obtained from Eq.(\ref{3.6}). For the evolution of the coherences, all the damping constants participate and the dynamical evolution is straightforwardly obtained from
\begin{eqnarray}
\frac{\partial \g{\rho}_{\alpha\beta}(t)}{\partial t}&=& -i\omega_{\alpha\beta}\g{\rho}_{\alpha\beta}(t)-\g{\Gamma}_{\alpha\beta\alpha\beta}\g{\rho}_{\alpha\beta}(t).
\label{3.9}
\end{eqnarray} 
For the population evolution, the transitions from low energy states to high energy states are usually neglected. Here, for the present molecular system, of which the coupled carbonyl stretches of ${\rm Rh}({\rm CO})_2({\rm C}_5{\rm H}_7{\rm O}_2)$ is a well-established example \cite{bb11}, some of these transitions constants associated with such transitions are orders of magnitude smaller than the other ones and will be neglected. However, for those pairs of states whose energy gap is small, they will be retained in our evaluation. Thus, with the numerical values of the eigenenergies obtained as functions of the mode-mode coupling interaction, we have 
\begin{eqnarray}
\g{\Gamma}=
\begin{pmatrix} 
\scr{0}\;&\;\scr{\g{\Gamma}_{\epsilon_1\epsilon_1\epsilon_2\epsilon_2}}\;&\;\scr{\g{\Gamma}_{\epsilon_1\epsilon_1\epsilon_3\epsilon_3}}\;&\;\scr{\g{\Gamma}_{\epsilon_1\epsilon_1\epsilon_4\epsilon_4}}\;&\;\scr{\g{\Gamma}_{\epsilon_1\epsilon_1\epsilon_5\epsilon_5}}\;&\;\scr{\g{\Gamma}_{\epsilon_1\epsilon_1\epsilon_6\epsilon_6}}\;\\
\scr{0}\;&\;\scr{\g{\Gamma}_{\epsilon_2\epsilon_2\epsilon_2\epsilon_2}}\;&\;\scr{\g{\Gamma}_{\epsilon_2\epsilon_2\epsilon_3\epsilon_3}}\;&\;\scr{\g{\Gamma}_{\epsilon_2\epsilon_2\epsilon_4\epsilon_4}}\;&\;\scr{\g{\Gamma}_{\epsilon_2\epsilon_2\epsilon_5\epsilon_5}}\;&\;\scr{\g{\Gamma}_{\epsilon_2\epsilon_2\epsilon_6\epsilon_6}}\;\\
\scr{0}\;&\;\scr{\g{\Gamma}_{\epsilon_3\epsilon_3\epsilon_2\epsilon_2}}\;&\;\scr{\g{\Gamma}_{\epsilon_3\epsilon_3\epsilon_3\epsilon_3}}\;&\;\scr{\g{\Gamma}_{\epsilon_3\epsilon_3\epsilon_4\epsilon_4}}\;&\;\scr{\g{\Gamma}_{\epsilon_3\epsilon_3\epsilon_5\epsilon_5}}\;&\;\scr{\g{\Gamma}_{\epsilon_3\epsilon_3\epsilon_6\epsilon_6}}\;\\
\scr{0}\;&\;\scr{0}\;&\;\scr{0}\;&\;\scr{\g{\Gamma}_{\epsilon_4\epsilon_4\epsilon_4\epsilon_4}}\;&\;\scr{\g{\Gamma}_{\epsilon_4\epsilon_4\epsilon_5\epsilon_5}}\;&\;\scr{\g{\Gamma}_{\epsilon_4\epsilon_4\epsilon_6\epsilon_6}}\;\\
\scr{0}\;&\;\scr{0}\;&\;\scr{0}\;&\;\scr{\g{\Gamma}_{\epsilon_5\epsilon_5\epsilon_4\epsilon_4}}\;&\;\scr{\g{\Gamma}_{\epsilon_5\epsilon_5\epsilon_5\epsilon_5}}\;&\;\scr{\g{\Gamma}_{\epsilon_5\epsilon_5\epsilon_6\epsilon_6}}\;\\
\scr{0}\;&\;\scr{0}\;&\;\scr{0}\;&\;\scr{0}\;&\;\scr{\g{\Gamma}_{\epsilon_6\epsilon_6\epsilon_5\epsilon_5}}\;&\;\scr{\g{\Gamma}_{\epsilon_6\epsilon_6\epsilon_6\epsilon_6}}\;\\
\end{pmatrix}
\label{3.10}
\end{eqnarray}  
and the equation of motion of the population takes the form
\begin{eqnarray}
\frac{\partial \g{\rho}_{\alpha\alpha}(t)}{\partial t}&=& -\sum_{\nu}\g{\Gamma}_{\alpha\alpha\nu\nu}\g{\rho}_{\nu\nu}(t).
\label{3.11}
\end{eqnarray}  
Finally, by identifying the integral representation of the density matrix with the definition of the population evolution Liouvillian, we have
\begin{eqnarray}
\g{\rho}_{\alpha\alpha}(t)=\frac{1}{2\pi i}\int_{-\infty+i\epsilon}^{\infty+i\epsilon}ds\, \e^{st}\bigl[s\g{I}+\g{\Gamma}\bigr]^{-1}_{\alpha\alpha\nu\nu}\g{\rho}_{\nu\nu}(t_0)
=\sum_{\nu}\g{G}_{\alpha\alpha\nu\nu}(t-t_0)\g{\rho}_{\nu\nu}(t_0),
\label{3.12}
\end{eqnarray} 
and the evolution Liouvillians in the population subspace are deduced. They correspond to $\g{G}_{\epsilon_1\epsilon_1\epsilon_1\epsilon_1}(t)=1$ for the ground state and
\begin{eqnarray}
\g{G}_{\epsilon_2\epsilon_2\epsilon_2\epsilon_2}(t)&=&\scr{\g{\Xi}}_{-+}\e^{\scr{\g{\lambda}_+}t}+ \scr{\g{\Xi}}_{+-}\e^{\scr{\g{\lambda}_-}t}       \nonumber\\
\g{G}_{\epsilon_3\epsilon_3\epsilon_3\epsilon_3}(t)&=&\scr{\g{\Xi}}_{+-}\e^{\scr{\g{\lambda}_+}t}+ \scr{\g{\Xi}}_{-+}\e^{\scr{\g{\lambda}_-}t}                  
\label{3.13}
\end{eqnarray} 
for the sole excited state population evolution participating in the 4WM photon-echo process considered here. Also, the notations
$\scr{\g{\lambda}_{\pm}}=-\frac{1}{2}( \scr{\g{\Gamma}_{\epsilon_2\epsilon_2\epsilon_2\epsilon_2}}+\scr{\g{\Gamma}_{\epsilon_3\epsilon_3\epsilon_3\epsilon_3}})
\pm \frac{1}{2} \scr{\g{\Delta}}$, $\scr{\g{\Xi}}_{-+}=(\scr{\g{\Delta}}-\scr{\g{\Gamma}_{\epsilon_2\epsilon_2\epsilon_2\epsilon_2}}+\scr{\g{\Gamma}_{\epsilon_3\epsilon_3\epsilon_3\epsilon_3}})/2\scr{\g{\Delta}}$ and
$\scr{\g{\Xi}}_{+-}=(\scr{\g{\Delta}}+\scr{\g{\Gamma}_{\epsilon_2\epsilon_2\epsilon_2\epsilon_2}}-\scr{\g{\Gamma}_{\epsilon_3\epsilon_3\epsilon_3\epsilon_3}})/2\scr{\g{\Delta}}$ have been introduced conjointly with the additional quantity
$\scr{\g{\Delta}}=\bigl[\scr{\g{\Gamma}_{\epsilon_2\epsilon_2\epsilon_2\epsilon_2}^2}-2\scr{\g{\Gamma}_{\epsilon_3\epsilon_3\epsilon_3\epsilon_3}}\scr{\g{\Gamma}_{\epsilon_2\epsilon_2\epsilon_2\epsilon_2}}+\scr{\g{\Gamma}_{\epsilon_3\epsilon_3\epsilon_3\epsilon_3}^2}+4\scr{\g{\Gamma}_{\epsilon_3\epsilon_3\epsilon_2\epsilon_2}}\scr{\g{\Gamma}_{\epsilon_2\epsilon_2\epsilon_3\epsilon_3}}\bigr]^{\frac{1}{2}}$. \\

To conclude this section, we shall discuss the physical meanings of the Liouvillian pathways participating in the 4WM photon-echo process according to the the phase-matched conditions determined by the geometry of the experiment, the energetic and temporal structures of the laser excitations and the usual restriction resulting from the rotating wave approximation (RWA). The first chronological field ordering corresponds to interactions with field components $-\vec{k}_a\rightarrow+\vec{k}_{b(c)}\rightarrow+\vec{k}_{c(b)}$ because of the field pulse overlapping. The second field ordering is given by $+\vec{k}_{b(c)}\rightarrow-\vec{k}_a\rightarrow+\vec{k}_{c(b)}$ and, finally, the last field ordering corresponds to $+\vec{k}_{b(c)}\rightarrow+\vec{k}_{c(b)}\rightarrow-\vec{k}_a$. Notice that for descriptions restricted to two- and three-level systems, this last field combination do not contribute. All these pathways are described in Appendix B. For each of them, their corresponding constants, say $A_{n,r,q,p}$, $B_{n,r,q}$, and $C_{n,r}$ as well as $\g{R}_{n,\ep i\ep j}(\tau_1,\tau_2,\tau_3,t)$ given in Eq.(\ref{2.7}), will be evaluated from the various matrix elements of the quantity $\langle \epsilon_i\vert\g{G}(t-\tau_3)\g{L}_{\rm v}(\tau_3)\g{G}(\tau_3-\tau_2)\g{L}_{\rm v}(\tau_2)\g{G}(\tau_2-\tau_1) \g{L}_{\rm v}(\tau_1)\g{\rho}(t_0)\vert\epsilon_j\rangle$ built from Eqs.(\ref{3.9}) and (\ref{3.13}). \\

\section{2D-infrared spectra: a detailed analysis} 

This Section is devoted to a detailed analysis of the 2D-spectra obtained by performing the double Fourier transformation of the usual 4WM photon-echo signal acquired by heterodyne detection in the direction $-\vec{k}_a+\vec{k}_b+\vec{k}_c$, as described previously. The first Fourier transform is performed with respect to the delay time between the laser field pulses, chosen as fields $a$ and $b$ here, that is $\tau=T_a-T_b=T_a$ due to our convention. The second transform is done with respect to the detection time $t$ as indicated in relation (\ref{2.16}). Also, we will discuss separately the real and the imaginary parts of $I_{4WM-pe}(\omega_{\tau},\omega_t)$, the complex 2D-spectrum of the 4WM photon-echo signal. They can be extracted by a convenient choice of the phase of the local field oscillator. Because the structure of these 2D-spectra is quite intricate, we divide the numerical simulations in two parts. In the first one, we will intent to give a comprehensive analysis of the various contributions participating in the Fourier spectra. Next, from this analysis, we will define appropriate conditions to obtain more informative experimental conditions to analyze the vibrational structure of the molecular systems. Finally, we have to point out that all the frequency dependences simulated here, are presented in arbitrary units. However, their relative magnitudes in the different figures are exact because the same multiplicative constant has been used throughout. This means that the vertical scale is the same for all the figures. The limited range of frequency used in our simulation is just a matter of convenience. It can be extended to include higher excited states, because all the states are taken into account in the dynamical evolution.

\subsection{\large Influence of the  overtones and combination states of higher energies}
The advantage of introducing a nonperturbative treatment of the mode-mode coupling and anharmonicity lies in having an exact resonance vibrational structure, that is, the exact vibrational energies and broadenings induced by the various transition and dephasing processes mixed by the mode-mode coupling interaction. Notice that except for those pathways whose contributions can be neglected because their dipole moments are order of magnitude smaller, all the other ones are taken into account exactly. Besides, since these experiments are usually performed with laser pulses in the femtosecond range, the spectral broadenings of the laser pulses are quite large and the 2D-spectra are strongly influenced by the overtones and combination states of higher energies, as we will see below. As indicated previously, heterodyne detection is introduced to get physical insights on the real and imaginary parts of the third-order polarization underlying the photon-echo signal. This is why the real and imaginary parts of the signal intensity  $I_{4WM-pe}(\omega_{\tau},\omega_t)$ will be analyzed separately.\\

To perform numerical simulations, we must introduce some specific values of the physical quantities required to characterize the excitation sources, as well as the vibrational molecular system. Concerning the excitation laser pulses, the probing times are set to be $T_b=T_c={\rm 0}\,{\rm ps}$. The laser bandwidths are chosen identical and correspond to $\gamma_a=\gamma_b=\gamma_c=60\,{\rm cm}^{-1}$ corresponding to $555\,{\rm fs}$. The excitation laser frequencies will be the same for the three beams. However, because the numerical simulations show different excitation conditions, their values will be indicated separately in the captions of the figures for each simulation. Next, consider the vibrational system.  Here, we express the values of the relaxation and dephasing constants in terms of the three parameters $a=3\,{\rm cm}^{-1},\; b=3\,{\rm cm}^{-1}$ and $c=5\,{\rm cm}^{-1}$. Therefore, the relaxation constants are $\Gamma_{1111}=0,\;\Gamma_{2222}=a,\;\Gamma_{3333}=b,\;\Gamma_{4444}=2a,\;\Gamma_{5555}=a+b,\;\Gamma_{6666}=2b$. With the pure dephasing constant $c$ assumed identical for all the states, all the dephasing constants $\Gamma_{ijij}$ are readily obtained. Moreover, we require the transition constants to satisfy the summation rule $\Gamma_{jjjj}=-\sum_{i\not=j}\Gamma_{iijj}$. Thus, we get $\Gamma_{11nn}=-\Gamma_{nnnn}\;{\rm if}\;n=1,2$, $\Gamma_{2244}=\Gamma_{3355}=\Gamma_{1122},\;\Gamma_{2255}=\Gamma_{3366}=\Gamma_{1133}$ and $\Gamma_{1144}=-\Gamma_{4444}-\Gamma_{2244}$, $\Gamma_{1155}=-\Gamma_{5555}-\Gamma_{2255}-\Gamma_{3355}$, and finally $\Gamma_{1166}=-\Gamma_{6666}-\Gamma_{3366}$.
As mentioned earlier, all the other relaxation, dephasing and transition constants are set to zero. Besides, the frequencies of the vibrational harmonic modes are $\omega_1=2015\,{\rm cm}^{-1}$ and $\omega_2=2084\,{\rm cm}^{-1}$. In addition, since for a vibrational mode, the transition dipole moments are usually assumed to satisfy the harmonic scaling law \cite{bb05,bb34}, the dipole matrix elements associated with two sequential excitations are related, for the fundamental and overtone transitions, by the expressions $\langle 01\vert\vec{\g{\mu}}\vert 02\rangle=\sqrt{2}\langle 00\vert\vec{\g{\mu}}\vert 01\rangle$ and $\langle 10\vert\vec{\g{\mu}}\vert 20\rangle=\sqrt{2}\langle 00\vert\vec{\g{\mu}}\vert 10\rangle$. Also, we assume the dipole moments real with the following values $\mu_{12}=\mu_{35}=1$, $\mu_{13}=\mu_{25}=1.2$, $\mu_{24}=\sqrt{2}$, and $\mu_{36}=\mu_{25}=1.2\sqrt{2}$. All the other dipole moments equal zero. 
Notice that an analogous scaling law can be found for vibrational population and phase relaxations \cite{bb50,bb51}. \\

\begin{figure}[!ht]
\centering
\includegraphics[clip,scale=0.82]{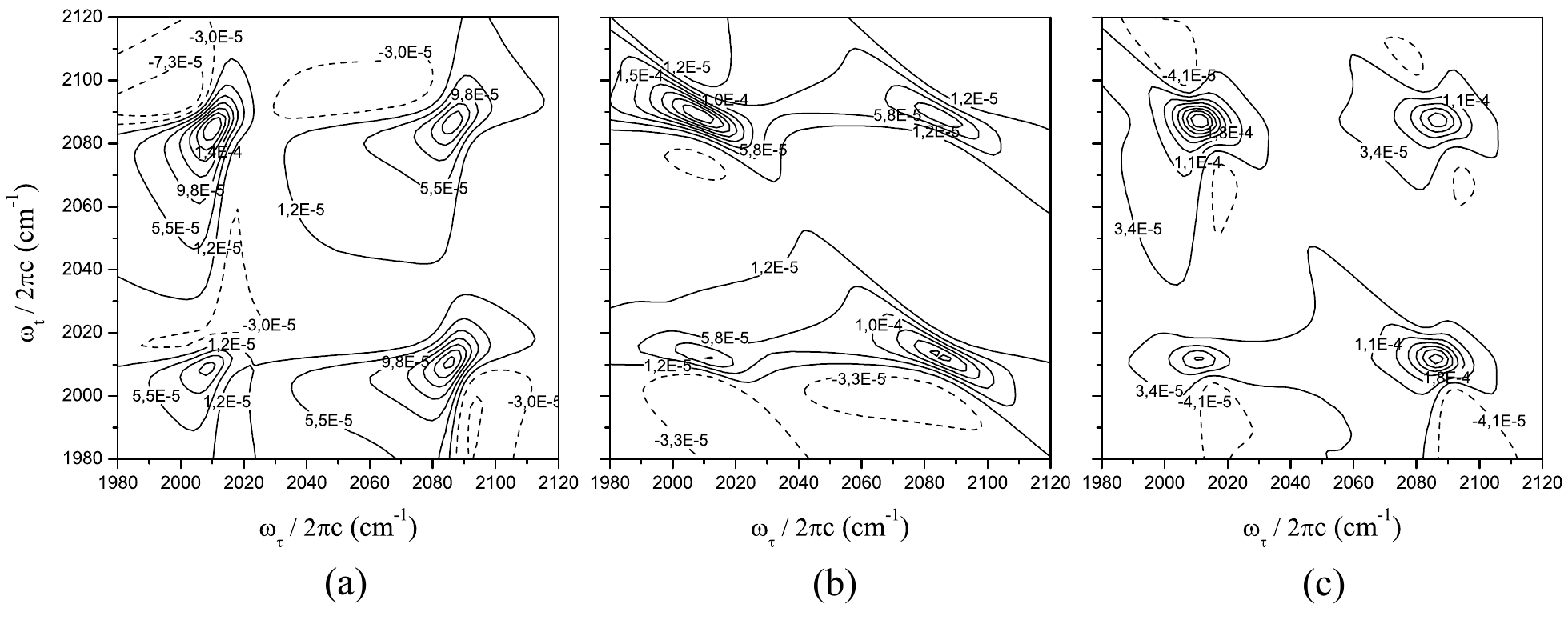}
\caption{2D-Fourier-transformed spectra of the 4WM photon-echo signal obtained by heterodyne detection in the direction $-\vec{k}_a+\vec{k}_b+\vec{k}_c$. Only the real parts are analyzed in this Figure. According to the types of chronological field ordering, the contributions from different pathways are presented in different panels: (a) the real part of the rephasing contribution, (b) the real part of the nonrephasing contribution, and (c) the real part of the total signal. The parameters of mode-anharmonicities are $u_{111}=u_{222}=10\,{\rm cm}^{-1}$ and the mode-mode coupling constants are $u_{12}=16\,{\rm cm}^{-1},\;u_{112}=u_{122}=4\,{\rm cm}^{-1}$. The laser frequencies are chosen all equal to $\omega_a=\omega_b=\omega_c=2015.5\,{\rm cm}^{-1}$. Notice that from this Figure onward, the positive contour levels are in solid lines and the negative contour levels are in dashed lines.}
\label{fig2}
\end{figure}

With these physical parameters, the numerical simulations can now be readily performed. In all the figures, the contour levels of positive values are shown in solid lines, while the contour levels of negative values are shown in dashed lines. First, we present in Fig.\ref{fig2} the two-dimensional frequency dependence of 
$I_{4WM-pe}(\omega_{\tau},\omega_{t})$. In the left most panel, Panel (a), we present the real part of the rephasing contribution corresponding to a process where field $a$ interacts first. We observed two diagonal peaks associated with the vibrational modes 1 and 2 located, for both coordinates, at the same eigenfrequencies $\omega_t\approx\omega_{\tau}\approx\epsilon_2-\epsilon_1$ for the first peak and $\omega_t\approx\omega_{\tau}\approx\epsilon_3-\epsilon_1$ for the second peak. Since we are concerned with the spectra of nonlinear optical responses, the positions of the resonances with respect to the eigenfrequency transitions of the vibrational modes can be slightly modified due to the influence of the other neighboring and overlapping resonances existing in any nonlinear signal structure. 
The other two resonances correspond to the cross-peaks and are located at the mixed eigenfrequencies $\left(\omega_t,\omega_{\tau}\right)\approx\left(\epsilon_3-\epsilon_1,\epsilon_2-\epsilon_1\right)$  for the upper-left peak and $\left(\omega_t,\omega_{\tau}\right)\approx\left(\epsilon_2-\epsilon_1,\epsilon_3-\epsilon_1\right)$ for the lower-right peak.  Notice that in the rephasing case, all the peak heights have approximately the same magnitude. Panel (b) associates with the non-rephasing contribution. The location of the resonances are similar to case (a), but the heights are greater by a factor of 2 for the cross peaks than for the diagonal peaks. Also, it is worthy of mentioning that these 2D-resonances have an egg-shaped structure with the main axis rotated by approximately $90 \degre$ between the rephasing and the non-rephasing contributions. In Panel (c), we show the total spectra. The two diagonal peaks and the two cross-peaks are reproduced. However, because of the compensation between the positive and negative values of the rephasing and non-rephasing contributions, the structures of the resonances are better localized and have more or less cylindrical symmetry. 
\begin{figure}[!ht]
\centering
\includegraphics[clip,scale=0.82]{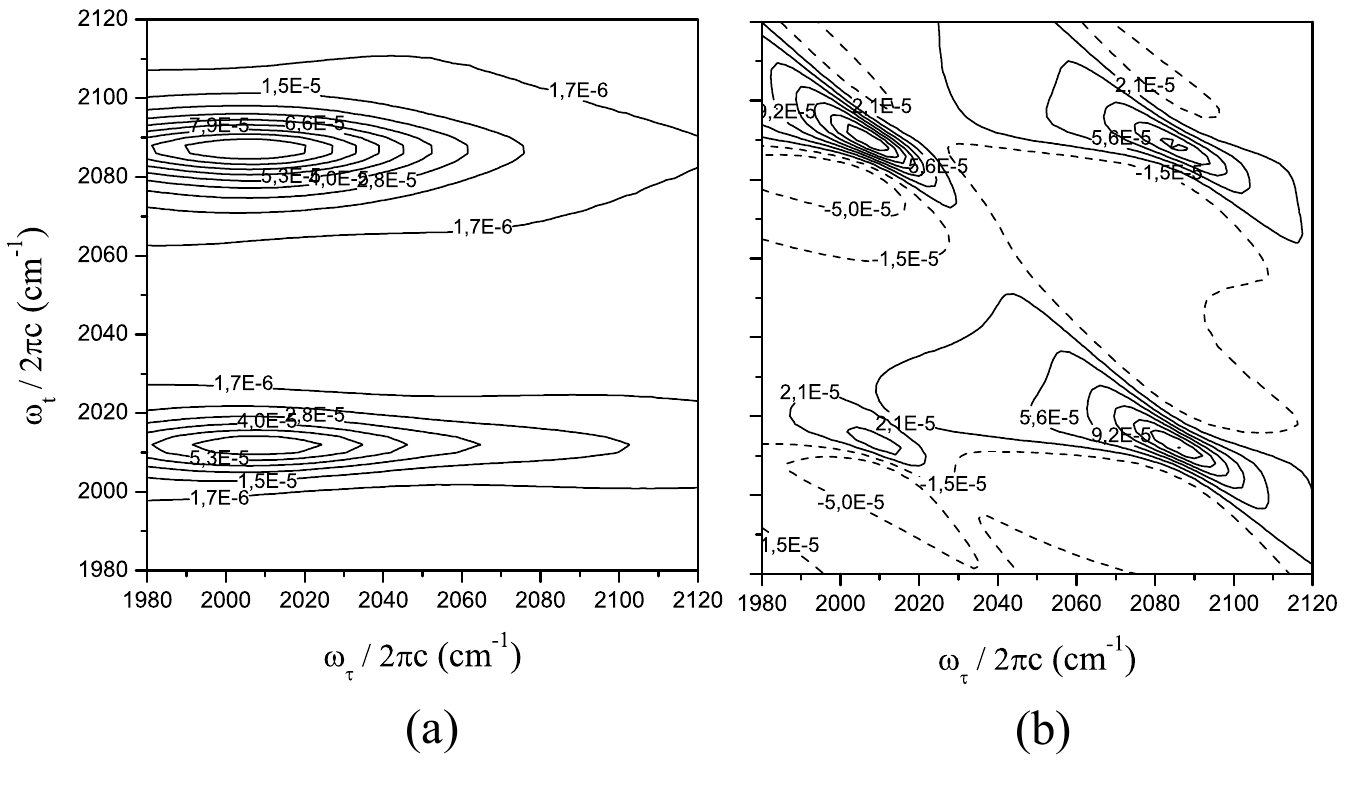}
\caption{Comparison of the real parts of the individual contributions participating in the non-rephasing terms of the 2D-Fourier spectra. The plot in Panel (a) represents the real part of the spectrum when field $a$ participates in the second interaction only, while (b) corresponds to the case where field $a$ participates solely to the third interaction. The physical parameters are identical to the ones used in the simulations shown in Fig.\ref{fig2}.}
\label{fig3}
\end{figure}

Next, in Fig.~\ref{fig3}, we present in two parts the contributions to the signal from nonrephasing pathways. Specifically, they refer respectively to whether the field $a$ participates in the second or the third system-field interaction. It has to be pointed out that for simple two- or three-level systems, the last contribution do not contribute to the 2D-spectrum. Panel (a) represents the real part of the contribution from the chronological field orderings $\vec{E}_{b(c)}\rightarrow\vec{E}_a\rightarrow\vec{E}_{c(b)}$. Only one diagonal peak and one cross-peak appear significantly and the orientation of the egg-shaped structure is parallel to the $\omega_{\tau}$ axis. Besides, the contribution is mainly positive. In contrast, contribution corresponding to the chronological ordering $\vec{E}_{b(c)}\rightarrow\vec{E}_{c(b)}\rightarrow\vec{E}_a$ is shown in Panel (b). Here we recover the four resonance structures. Moreover, by comparing Figs.3a and 3b with Fig.2b, we see that the orientation of the resonance shape in the non-rephasing case clearly originates from this last contribution.

\begin{figure}[!ht]
\centering
\includegraphics[clip,scale=0.82]{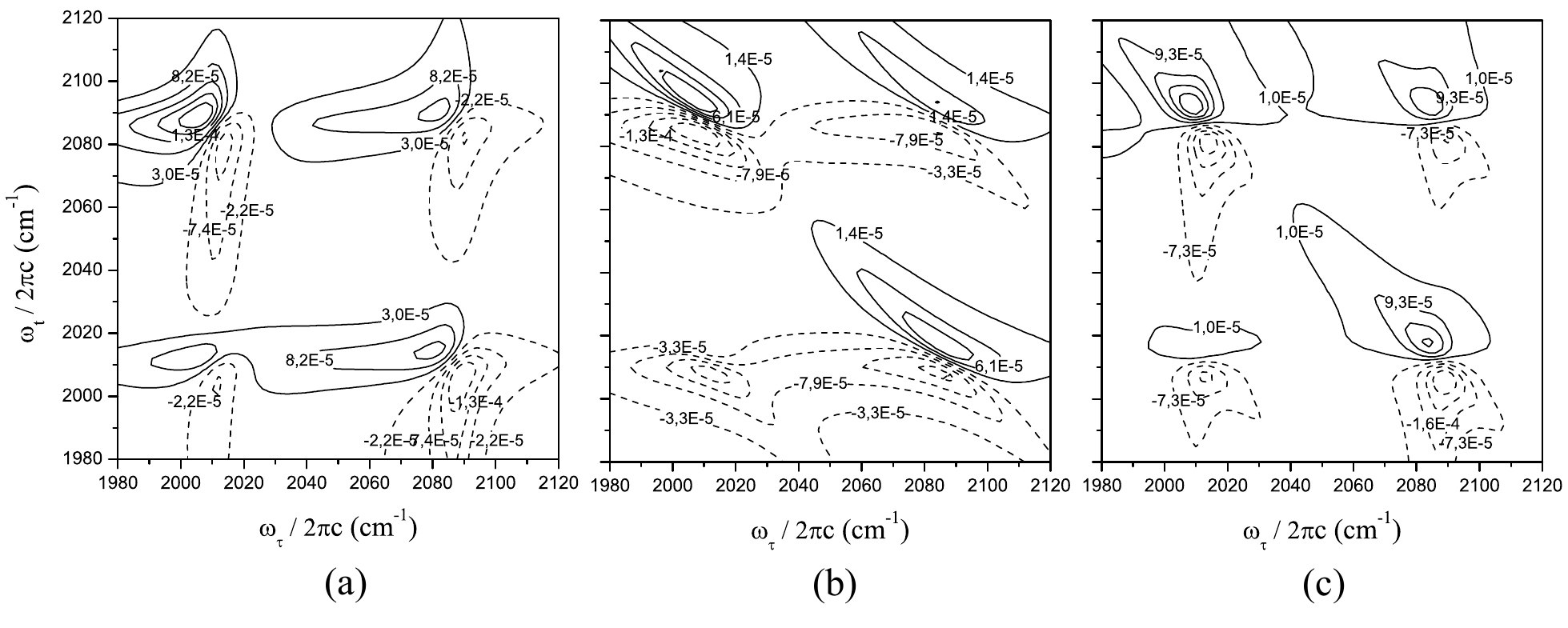}
\caption{The imaginary parts of the rephasing contribution (a), non-rephasing contribution (b) and total contribution (c) to the 2D spectra. We use the same values for the physical parameters as the ones introduced in Fig.\ref{fig2}.} 
\label{fig4}
\end{figure}
\begin{figure}[!ht]
\centering
\includegraphics[clip,scale=0.82]{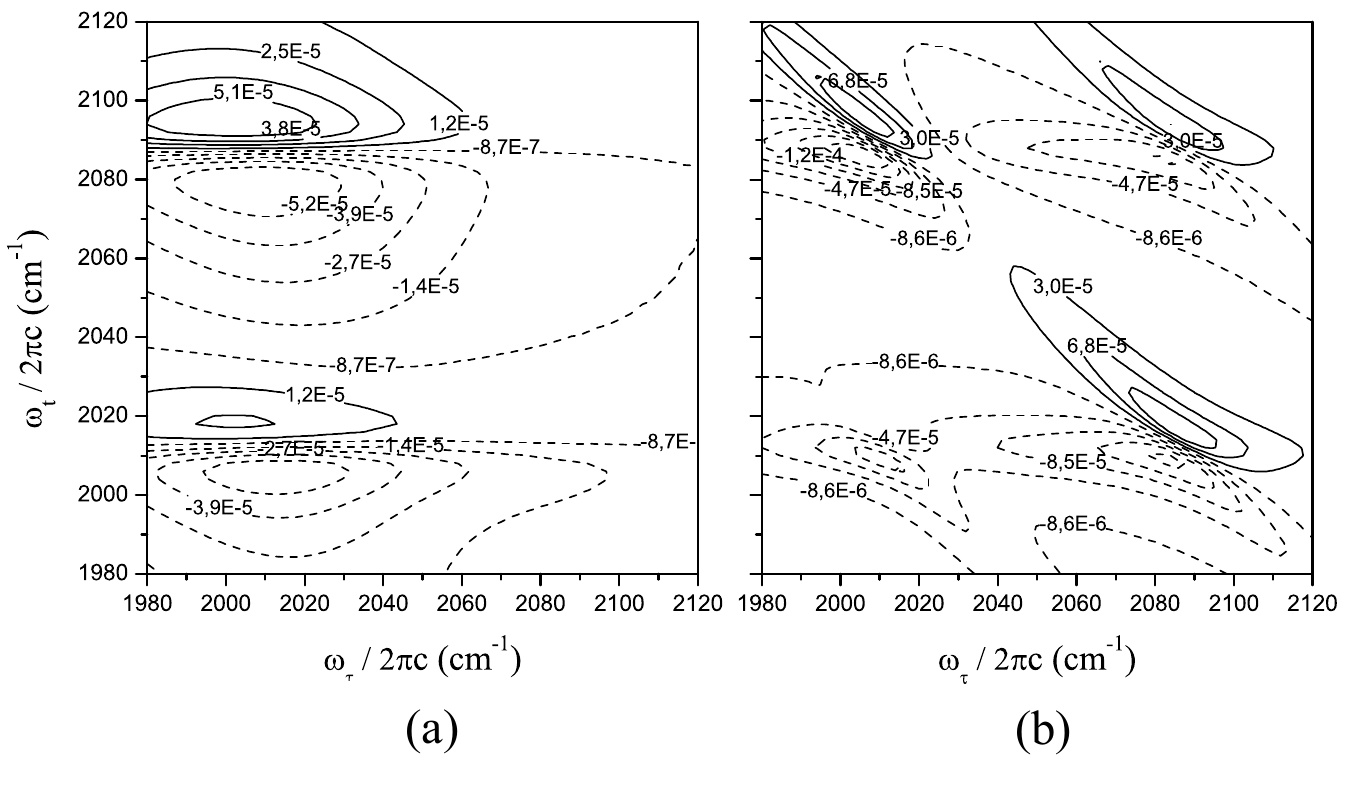}
\caption{Comparison of the imaginary parts of the individual contributions participating in the non-rephasing terms of the 2D-Fourier spectrum. In this figure, (a) represents the imaginary part of the spectrum when field $a$ participates to the second interaction only, while (b) corresponds to the case where field $a$ participates solely to the third interaction. The physical parameters are identical to the ones used in previous simulations shown in Fig.\ref{fig2}.}
\label{fig5}
\end{figure}

It is also necessary to discuss the imaginary parts of the spectra of (a) the rephasing contribution, (b) the non-rephasing contribution, and (c) total contribution to the 2D-spectrum. They are  shown in Fig.4. Following the same line in the discussion of the real parts, we start with the variations of the rephasing term shown in Panel (a). Again, the four resonance structures made of two diagonal peaks and two cross-peaks are observed. However, each resonance peak has a more butterfly-like shape, in contrast to the egg-shaped peaks in the real part spectrum.  
These typical features of the real and imaginary parts of the 2D spectrum can be related to the profiles of the imaginary and real parts of the linear polarization associated with the dispersion curve and absorption spectrum, respectively. Similar to what we saw in the real part of the signal, a rotation of the individual resonance structures is observed. Also, by summing up the rephasing and the nonrephasing contributions, there is a compensation of the positive and negative contributions which results in a sharper localization of the resonances, as can be seen in Fig.\ref{fig4}(c), where the imaginary part of the total 2D spectra is presented. Finally, in Fig.\ref{fig5}, we compare the two terms participating in the nonrephasing contribution and corresponding to the chronological orderings $\vec{E}_{b(c)}\rightarrow\vec{E}_a\rightarrow\vec{E}_{c(b)}$ in Panel (a) and $\vec{E}_{b(c)}\rightarrow\vec{E}_{c(b)}\rightarrow\vec{E}_a$ in Panel (b), respectively. Besides, we recover the preferential orientations of the main axis of these resonance structures observed in Fig.2, which are now broader than the ones observed in Figs.3a and 3b.

\subsection{\large Non-resonant excitation effects}

The final topic to discuss is the off-resonance effects. Even though the short laser pulses used in these experiments imply quite broad spectral distributions, off-resonant excitation conditions can be sometimes useful to disentangle some intricate resonance structure. While qualitatively these effects can be easily understood, quantitatively the problem is more intricate. To this end, we have done a series of numerical simulations to analyze the 2D-vibrational structure as a function of the off-resonance parameters. In these simulations, we have slightly modified the values of the coupling constants as indicated in the caption of Fig.6. Accordingly, the positions of the diagonal peaks and cross-peaks are shifted  as shown in the following figures. The resonance shapes are more or less egg-shaped or butterfly-shaped depending on whether we are testing the real or the imaginary part of the rephasing and non-rephasing contributions, respectively. This observation remains true far any laser excitation frequency in the full range of excitation laser frequency we have considered.
\begin{figure}[!ht]
\centering
\includegraphics[clip,scale=0.82]{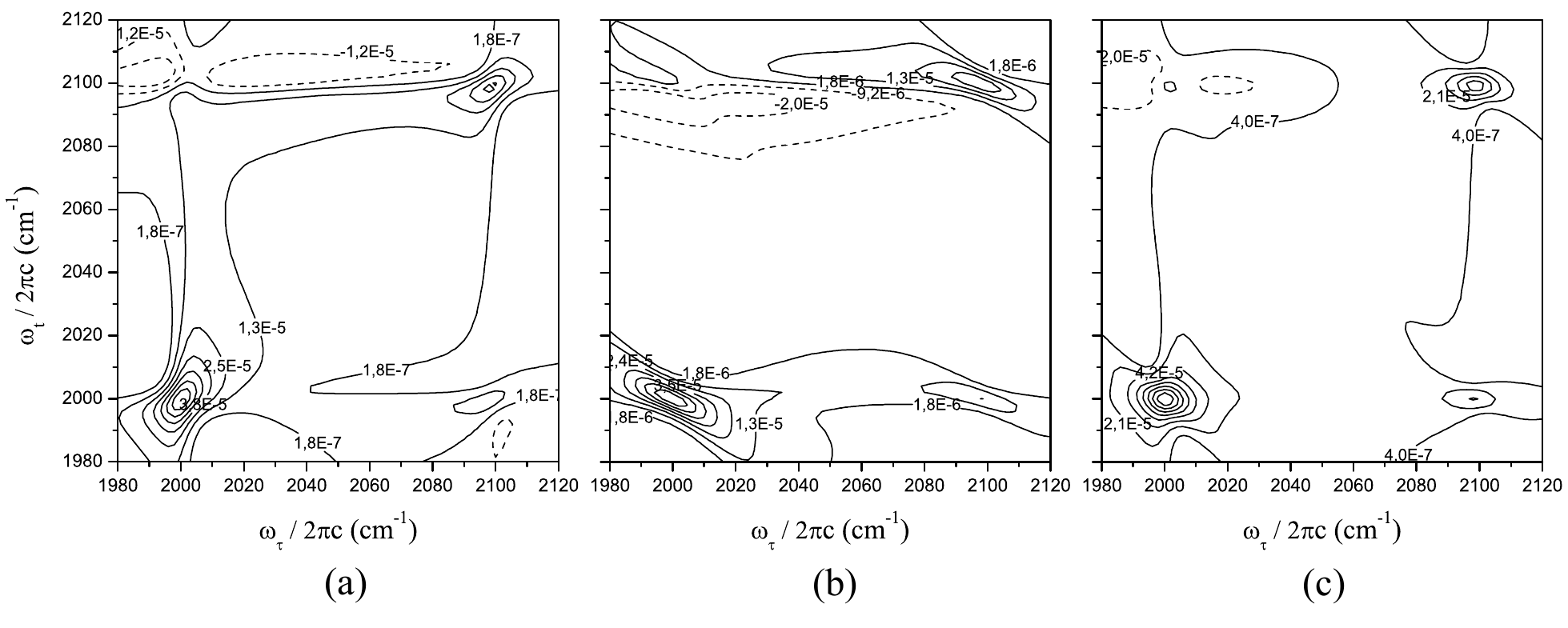}
\caption{2D-Fourier-transformed spectra of the heterodyned 4WM photon-echo signal. Again, different contributions are presented in separate panels: (a) the real rephasing contribution, (b) the real non-rephasing contribution and (c) for the real part of the total signal. Anharmonicity  and mode-mode coupling constants of the vibrational modes are now characterized by the constants $u_{111}=u_{222}=10\,{\rm cm}^{-1}\;u_{12}=36\,{\rm cm}^{-1},\;u_{112}=u_{122}=6\,{\rm cm}^{-1}$. All the laser frequencies are chosen equal to $\omega_a=\omega_b=\omega_c=2015.5\,{\rm cm}^{-1}$. As previously indicated, continuous lines stand for positive level contours and dashed lines for negative level contours.}
\label{fig6}
\end{figure}
Another feature that remains the same is the fact that the combination of the rephasing and nonrephasing contributions leads to a better localization of the vibrational resonance. 
\begin{figure}[!ht]
\centering
\includegraphics[clip,scale=0.82]{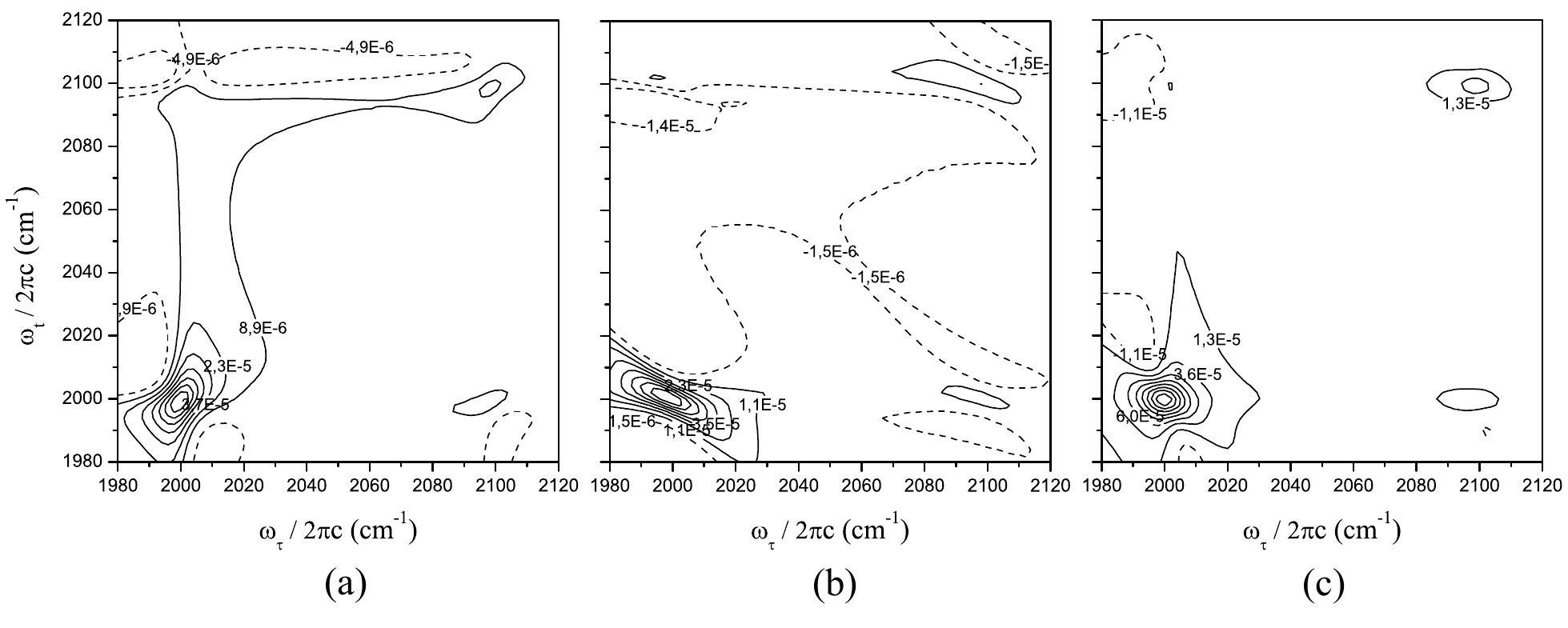}
\caption{The same kinds of spectra as in Fig.\ref{fig6} under a different excitation condition. Here, the laser frequencies are all the same and correspond to $\omega_a=\omega_b=\omega_c=2000.5\,{\rm cm}^{-1}$. In Panels (a), (b) and (c) the real rephasing contribution, the real non-rephasing contribution and the real part of the total signal, respectively, are presented.}
\label{fig7}
\end{figure}
In Figs.6 to 8, we plot the real parts of the (a) rephasing, (b) non-rephasing and (c) total contributions to the 2D-spectra obtained for different excitation laser frequencies: $\omega_a=\omega_b=\omega_c=2015.5\,{\rm cm}^{-1}$ in Fig.6, $\omega_a=\omega_b=\omega_c=2000.5\,{\rm cm}^{-1}$ in Fig. 7, and $\omega_a=\omega_b=\omega_c=1975.5\,{\rm cm}^{-1}$ in Fig. 8, respectively. Comparing Fig.2 and Fig.6, it can be noticed that while the cross-peaks are higher than the diagonal peaks in the Fig.2, the situation is reversed in Fig.6. This is due to the fact that by increasing the mode-mode coupling term, we shift the transition frequencies of the vibrational system. As a consequence, the laser excitation conditions $\omega_a=\omega_b=\omega_c=2015.5\,{\rm cm}^{-1}$ in Fig.2 are more resonant to the vibrational transition frequencies than in Fig.6. This phenomenon is more enhanced in Figs.7 and 8.
\begin{figure}[!ht]
\centering
\includegraphics[clip,scale=0.82]{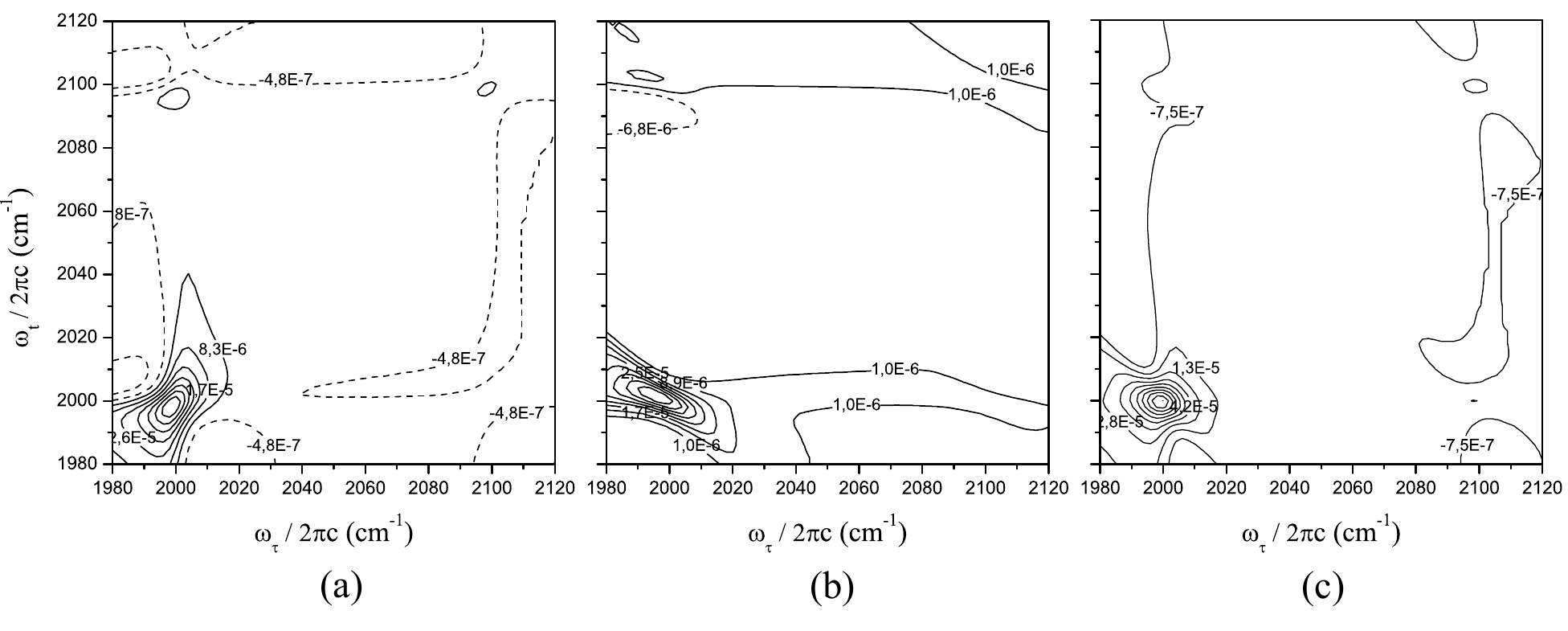}
\caption{Similar plots as in Figs.\ref{fig6} and \ref{fig7} with the excitation laser frequencies  $\omega_a=\omega_b=\omega_c=1975.5\,{\rm cm}^{-1}$.}
\label{fig8}
\end{figure}

\begin{figure}[!ht]
\centering
\includegraphics[clip,scale=0.82]{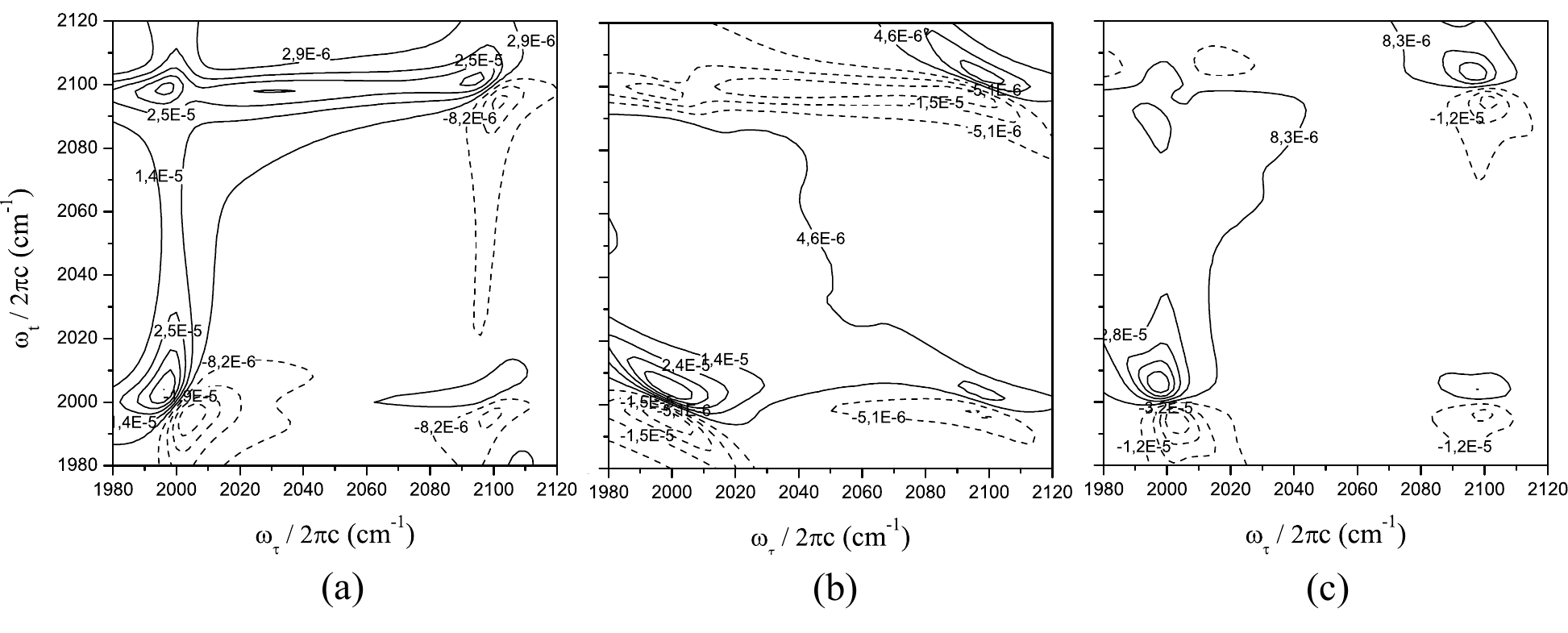}
\caption{The imaginary spectra of the case shown in Fig.\ref{fig6}. The laser excitation frequencies are $\omega_a=\omega_b=\omega_c=2015.5\,{\rm cm}^{-1}$.}
\label{fig9}
\end{figure}

\begin{figure}[!ht]
\centering
\includegraphics[clip,scale=0.82]{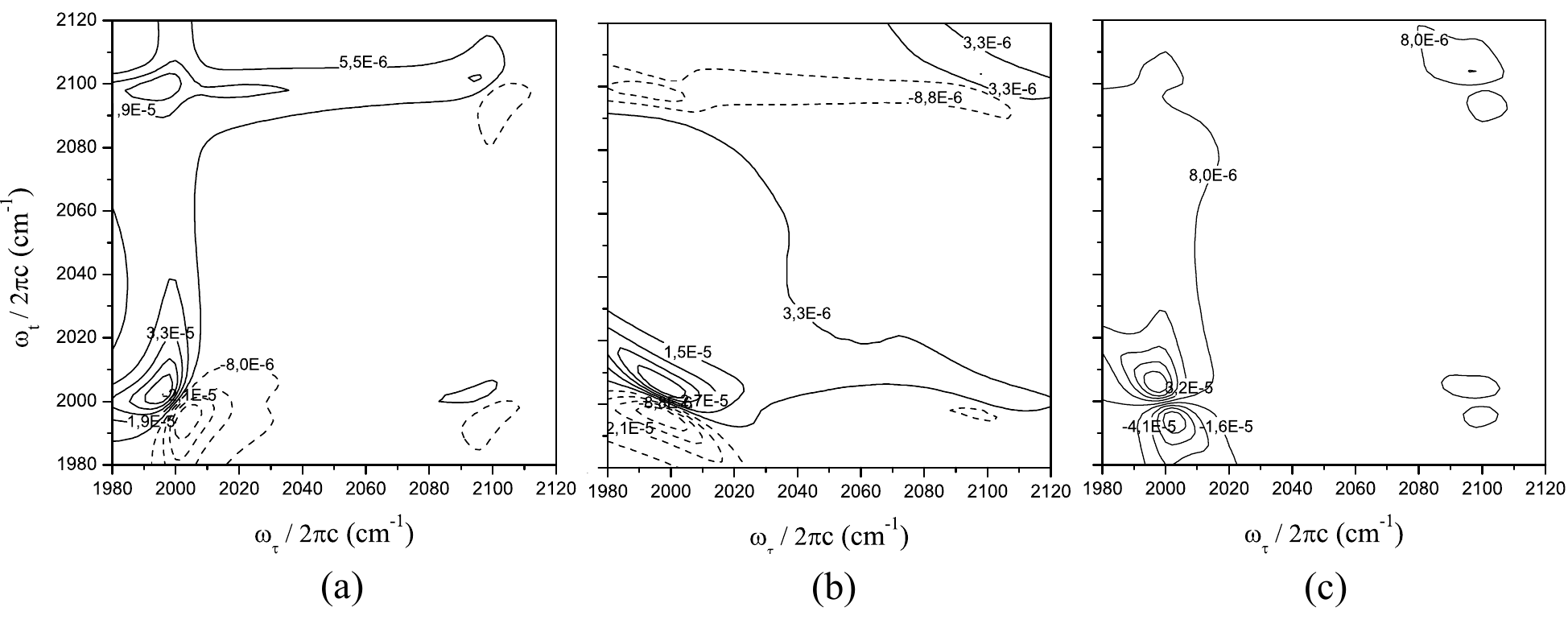}
\caption{The imaginary spectra of the case shown in Fig.\ref{fig7}. The laser excitation frequencies are $\omega_a=\omega_b=\omega_c=2000.5\,{\rm cm}^{-1}$.}
\label{fig10}
\end{figure}

\begin{figure}[!t]
\centering
\includegraphics[clip,scale=0.82]{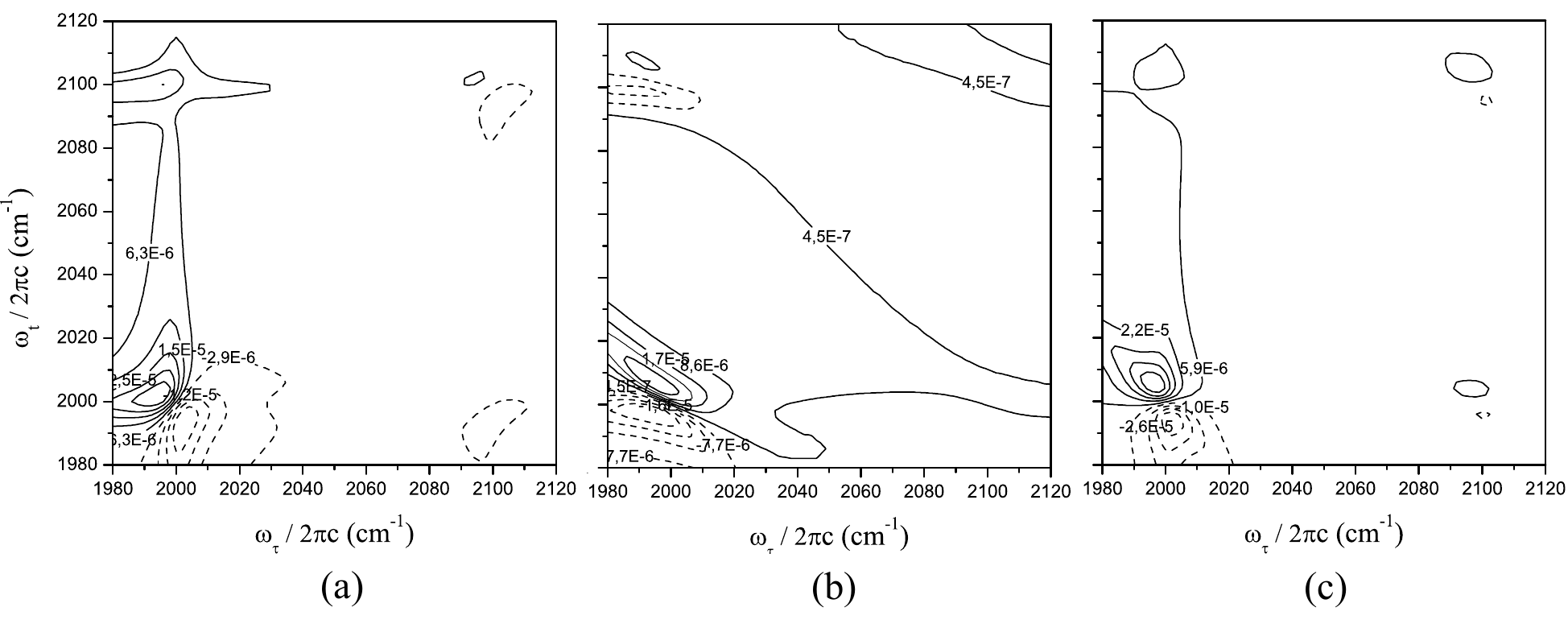}
\caption{The imaginary spectra of the case shown in Fig.\ref{fig8}. The laser excitation frequencies are $\omega_a=\omega_b=\omega_c=1975.5\,{\rm cm}^{-1}$.}
\label{fig11}
\end{figure}
With the mode-mode coupling parameters we used, after diagonalizing the system we find that the transition frequencies are $\omega_{\ep2\ep1}=1999.877\,{\rm cm}^{-1}$ and     $\omega_{\ep3\ep1}=2099.400\,{\rm cm}^{-1}$. Therefore, the off-resonance parameter  $\omega_{\ep2\ep1}-\omega_a$ has the values  $-15.623\,{\rm cm}^{-1}$ for Fig.6, $-0.623\,{\rm cm}^{-1}$ for Fig.7 and $24.377\,{\rm cm}^{-1}$ for Fig.8. The magnitudes of the lower resonance follow the variations of the off-resonance parameters.    
Similarly, for the second resonance, we have the values $83.900\,{\rm cm}^{-1}$ for Fig.6, $98.900\,{\rm cm}^{-1}$ for Fig.7 and $123.900\,{\rm cm}^{-1}$ for Fig.8.
Finally, in Figures 9, 10 and 11, the imaginary parts of the spectra with the same physical parameters are presented. The values of the parameters are listed in respective captions.

\section{Conclusion} 

In the present work, we have derived an analytical description which allows a detailed analysis of the 2D infrared spectra of a vibrational molecular system. Starting from a model of anharmonic coupled oscillators handled non-perturbatively and using heterodyne detection of the 4WM photon-echo signal, the various contributions from the rephasing and non-rephasing pathways to the 2D-spectrum have been evaluated. The non-rephasing term can itself be separated in two terms associated with specific chronological orderings of the excitation laser beams knowing that for simple two- or three-level systems, this last contribution disappears. 
The numerical simulations performed from the analytical expressions of the 4WM photon-echo signal, show characteristic shapes of the exact resonance structure involving a complete redistribution of the dipole moments as well as of the various relaxation and dephasing constants. We observe that summation of the rephasing and non-rephasing contributions leads to a more localized resonance. Also, with a convenient choice of the local field oscillator phase, we can select the real or the imaginary parts of the 2D-spectra having specific egg- or butterfly-shaped resonance whose shape along the line joining the minimum and maximum of the spectra can be related to the dispersion or absorption curves of the linear response function. Finally, the near-resonant excitation effects have been analyzed quantitatively. 

\begin{acknowledgments} 
This work was supported by the Academia Sinica and by the National Science Council of Taiwan (grant NSC 100-2113-M-001-006-MY2). One of us, (A.A. Villaeys), is indebted to Dr. K.K. Liang for his kind hospitality during his stay in Nankang. 
\end{acknowledgments} 

\appendix

\section{Triple-time integral}

We present the procedure to formally integrate the triple-time integral required to evaluate the photon-echo signal underlying the 2D-spectra. The first formal time-integration gives the expression
\begin{eqnarray}
\int_{-\infty}^{t_j}dt_i\mathcal{A}_r(t_i-T_r)\e^{Xt_i}=H(T_r-t_j)P_i[1]\e^{p_i[1]t_j}+H(t_j-T_r)\sum_{\alpha=1}^{2}P_s[\alpha]\e^{p_s[\alpha]t_j}
\label{A.1} 
\end{eqnarray}
with the following constants
\begin{center}
\scriptsize{
\begin{supertabular}{lll}
\hline
\hline
$P_i[1]=\sqrt{\gamma_r}\e^{-\gamma_rT_r}[X+\gamma_r]^{-1}$ &$p_i[1]=X+\gamma_r$ &              \\
\hline
$P_s[1]=\sqrt{\gamma_r}\e^{\gamma_rT_r}[X-\gamma_r]^{-1}$  &$p_s[1]=X-\gamma_r$  &               \\
\hline
$P_s[2]=\sqrt{\gamma_r}\e^{XT_r}[X+\gamma_r]^{-1}-\e^{XT_r}[X-\gamma_r]^{-1}$ &$p_s[2]=0$  &            \\
\hline
\hline
\end{supertabular}} 
\end{center} 
The second time-integration results into the form
{\small %
\begin{multline}
\int_{-\infty}^{t_k}dt_j\mathcal{A}_q(t_j-T_q)\e^{Yt_j}\int_{-\infty}^{t_j}dt_i\mathcal{A}_r(t_i-T_i)\e^{Xt_i}
=H(T_r-T_q)\bigl[H(T_q-t_k)D_{i-rsq}[1]\e^{d_{i-rsq}[1]t_k} \\
+H(t_k-T_q)H(T_r-t_k)D_{m-rsq}[1]\e^{d_{m-rsq}[1]t_k} +\sum_{\alpha=1}^{4}H(t_k-\tau_{s-rsq}[\alpha])D_{s-rsq}[\alpha]\e^{d_{s-rsq}[\alpha]t_k}\bigr]  \\
+H(T_q-T_r)\bigl[H(T_r-t_k)D_{i-qsr}[1]\e^{d_{i-qsr}[1]t_k}+H(t_k-T_r)H(T_q-t_k)\sum_{\alpha=1}^{2}D_{m-qsr}[\alpha]\e^{d_{m-qsr}[\alpha]t_k}     \\
+\sum_{\alpha=1}^{4}H(t_k-\tau_{s-qsr}[\alpha])D_{s-qsr}[\alpha]\e^{d_{s-qsr}[\alpha]t_k}\bigr]
\label{A.2} 
\end{multline}}
where the constants are given in the following tables and they stand for
\begin{center}
{\scriptsize
\begin{supertabular}{lll}
\hline
\hline
 $D_{i-rsq}[1]=\sqrt{\gamma_q}P_i[1]\e^{-\gamma_qT_q}\bigl[Y+\gamma_q+p_i[1]\bigr]^{-1}$  
&$d_{i-rsq}[1]=Y+\gamma_q+p_i[1]$ 
&                             \\
 $D_{m-rsq}[1]=\sqrt{\gamma_q}P_i[1]\e^{\gamma_qT_q}\bigl[Y-\gamma_q+p_i[1]\bigr]^{-1}$ 
&$d_{m-rsq}[1]=Y-\gamma_q+p_i[1]$   
&                             \\
 $D_{s-rsq}[1]=\sqrt{\gamma_q}P_i[1]\e^{YT_q+p_i[1]T_q}  \Bigl[\bigl[Y+\gamma_q+p_i[1]\bigr]^{-1}-\bigl[Y-\gamma_q+p_i[1]\bigr]^{-1}\Bigr]$
&$d_{s-rsq}[1]=0$
&$\tau_{s-rsq}[1]=T_q$        \\
 $D_{s-rsq}[2]=\sqrt{\gamma_q}P_s[1]\e^{\gamma_qT_q}\bigl[Y-\gamma_q+p_s[1]\bigr]^{-1}$
&$d_{s-rsq}[2]=Y-\gamma_q+p_s[1]$
&$\tau_{s-rsq}[2]=T_r$        \\
 $D_{s-rsq}[3]=\sqrt{\gamma_q}P_s[2]\e^{\gamma_qT_q}\bigl[Y-\gamma_q+p_s[2]\bigr]^{-1}$
&$d_{s-rsq}[3]=Y-\gamma_q+p_s[2]$
&$\tau_{s-rsq}[3]=T_r$        \\
$D_{s-rsq}[4]=\sqrt{\gamma_q}P_i[1]\e^{YT_r-\gamma_q(T_r-T_q)+p_i[1]T_r}\bigl[Y-\gamma_q+p_i[1]\bigr]^{-1}$
&$d_{s-rsq}[4]=0$
&$\tau_{s-rsq}[4]=T_r$        \\
 $\hskip 1.5truecm -\sqrt{\gamma_q}\sum_{\alpha=1}^{2}P_s[\alpha]\e^{YT_r-\gamma_q(T_r-T_q)+p_s[\alpha]T_r}\bigl[Y-\gamma_q+p_s[\alpha]\bigr]^{-1}$
&        
&              \\
\hline
\hline
\end{supertabular}}
\end{center}
if the condition $T_r>T_q$ is satisfied and 
\begin{center}
{\scriptsize
\begin{supertabular}{lll}
\hline
\hline
 $D_{i-qsr}[1]=\sqrt{\gamma_q}P_i[1]\e^{-\gamma_qT_q}\bigl[Y+\gamma_q+p_i[1]\bigr]^{-1}$
&$d_{i-qsr}[1]=Y+\gamma_q+p_i[1]$
&                             \\
 $D_{m-qsr}[1]=\sqrt{\gamma_q}P_s[1]\e^{-\gamma_qT_q}\bigl[Y+\gamma_q+p_s[1]\bigr]^{-1}$
&$d_{m-qsr}[1]=Y+\gamma_q+p_s[1]$
&                             \\
 $D_{m-qsr}[2]=\sqrt{\gamma_q}P_s[2]\e^{-\gamma_qT_q}\bigl[Y+\gamma_q+p_s[2]\bigr]^{-1}$
&$d_{m-qsr}[2]=Y+\gamma_q+p_s[2]$
&                             \\
 $D_{s-qsr}[1]=\sqrt{\gamma_q}P_i[1]\e^{YT_r+\gamma_q(T_r-T_q)+p_i[1]T_r}\bigl[Y+\gamma_q+p_i[1]\bigr]^{-1}$
&$d_{s-qsr}[1]=0$
&$\tau_{s-qsr}[1]=T_r$        \\
$\hskip 1.5truecm +\sqrt{\gamma_q}\sum_{\beta=1}^{2}P_s[\beta]\e^{YT_r+\gamma_q(T_r-T_q)+p_s[\beta]T_r}\bigl[-Y-\gamma_q-p_s[\beta]\bigr]^{-1}$ 
&
&         \\
 $D_{s-qsr}[2]=\sqrt{\gamma_q}P_s[1]\e^{\gamma_qT_q}\bigl[Y-\gamma_q+p_s[1]\bigr]^{-1}$
&$d_{s-qsr}[2]=Y-\gamma_q+p_s[1]$
&$\tau_{s-qsr}[2]=T_q$        \\
 $D_{s-qsr}[3]=\sqrt{\gamma_q}P_s[2]\e^{\gamma_qT_q}\bigl[Y-\gamma_q+p_s[2]\bigr]^{-1}$
&$d_{s-qsr}[3]=Y-\gamma_q+p_s[2]$
&$\tau_{s-qsr}[3]=T_q$        \\
$D_{s-qsr}[4]=\sqrt{\gamma_q}\sum_{\beta=1}^{2}P_s[\beta]\e^{YT_q+p_s[\beta]T_q}$
&$d_{s-qsr}[4]=0$
&$\tau_{s-qsr}[4]=T_q$        \\
$\hskip 3truecm \times\Bigl[\bigl[Y+\gamma_q+p_s[\beta]\bigr]^{-1}-\bigl[Y-\gamma_q+p_s[\beta]\bigr]^{-1}    \Bigr]$
&
&        \\
\hline
\hline
\end{supertabular}} 
\end{center} 
in the opposite situation where $T_q>T_r$.

\newpage\rotatebox{90}{\begin{minipage}{\textheight}
Now, we come to the third time-integration. Here, six different chronological ordering participate in the process. The formal result takes the form
{\tiny
\begin{eqnarray}
\int_{-\infty}^{t}&d&t_k\mathcal{A}_p(t_k-T_p)\e^{Zt_k}\int_{-\infty}^{t_k}dt_j\mathcal{A}_q(t_j-T_q)\e^{Yt_j}\int_{-\infty}^{t_j}dt_i
\mathcal{A}_r(t_i-T_i)\e^{Xt_i}                                                                                                                    \nonumber\\
&=&H(T_p-T_q)H(T_q-T_r)\Bigg\{H(\tau_{i-pqr}-t)F_{i-pqr}[1]\e^{f_{i-pqr}[1]t}+\sum_{\alpha=1}^6 H(t-\tau_{m-pqr-1}[\alpha])H(\tau_{m-pqr-2}[\alpha]-t) F_{m-pqr}[\alpha]\e^{f_{m-pqr}[\alpha]t}                                                                                                           \nonumber\\
&&\hskip 3.7truecm+\sum_{\alpha=1}^7 H(t-\tau_{s-pqr}[\alpha])F_{s-pqr}[\alpha]\e^{f_{s-pqr}[\alpha]t}\Bigg\}                                        \nonumber\\
&&+H(T_q-T_p)H(T_p-T_r)\Bigg\{H(\tau_{i-qpr}-t)F_{i-qpr}[1]\e^{f_{i-qpr}[1]t}+\sum_{\alpha=1}^4 H(t-\tau_{m-qpr-1}[\alpha])H(\tau_{m-qpr-2}[\alpha]-t) F_{m-qpr}[\alpha]\e^{f_{m-qpr}[\alpha]t}                                                                                                           \nonumber\\
&&\hskip 3.7truecm+\sum_{\alpha=1}^8 H(t-\tau_{s-qpr}[\alpha])F_{s-qpr}[\alpha]\e^{f_{s-qpr}[\alpha]t}\Bigg\}                                        \nonumber\\
&&+H(T_p-T_r)H(T_r-T_q)\Bigg\{H(\tau_{i-prq}-t)F_{i-prq}[1]\e^{f_{i-prq}[1]t}+\sum_{\alpha=1}^5 H(t-\tau_{m-prq-1}[\alpha])H(\tau_{m-prq-2}[\alpha]-t) F_{m-prq}[\alpha]\e^{f_{m-prq}[\alpha]t}                                                                                                           \nonumber\\
&&\hskip 3.7truecm+\sum_{\alpha=1}^7 H(t-\tau_{s-prq}[\alpha])F_{s-prq}[\alpha]\e^{f_{s-prq}[\alpha]t}\Bigg\}                                        \nonumber\\
&&+H(T_r-T_p)H(T_p-T_q)\Bigg\{H(\tau_{i-rpq}-t)F_{i-rpq}[1]\e^{f_{i-rpq}[1]t}+\sum_{\alpha=1}^3 H(t-\tau_{m-rpq-1}[\alpha])H(\tau_{m-rpq-2}[\alpha]-t) F_{m-rpq}[\alpha]\e^{f_{m-rpq}[\alpha]t}                                                                                                           \nonumber\\
&&\hskip 3.7truecm+\sum_{\alpha=1}^7 H(t-\tau_{s-rpq}[\alpha])F_{s-rpq}[\alpha]\e^{f_{s-rpq}[\alpha]t}\Bigg\}                                        \nonumber\\
&&+H(T_q-T_r)H(T_r-T_p)\Bigg\{H(\tau_{i-qrp}-t)F_{i-qrp}[1]\e^{f_{i-qrp}[1]t}+\sum_{\alpha=1}^3 H(t-\tau_{m-qrp-1}[\alpha])H(\tau_{m-qrp-2}[\alpha]-t) F_{m-qrp}[\alpha]\e^{f_{m-qrp}[\alpha]t}                                                                                                           \nonumber\\
&&\hskip 3.7truecm+\sum_{\alpha=1}^7 H(t-\tau_{s-qrp}[\alpha])F_{s-qrp}[\alpha]\e^{f_{s-qrp}[\alpha]t}\Bigg\}                                        \nonumber\\
&&+H(T_r-T_q)H(T_q-T_p)\Bigg\{H(\tau_{i-rqp}-t)F_{i-rqp}[1]\e^{f_{i-rqp}[1]t}+\sum_{\alpha=1}^2 H(t-\tau_{m-rqp-1}[\alpha])H(\tau_{m-rqp-2}[\alpha]-t) F_{m-rqp}[\alpha]\e^{f_{m-rqp}[\alpha]t}                                                                                                           \nonumber\\
&&\hskip 3.7truecm+\sum_{\alpha=1}^7 H(t-\tau_{s-rqp}[\alpha])F_{s-rqp}[\alpha]\e^{f_{s-rqp}[\alpha]t}\Bigg\}                                     
\label{A.3} 
\end{eqnarray}}
and all the constants are listed below according to their particular ordering indicated by the subscripts $pqr$, $qpr$, $prq$, $rpq$, $qrp$ and $rqp$ successively.
\end{minipage}}\newpage
\newpage\rotatebox{90}{\begin{minipage}{\textheight}
First case: $T_p>T_q>T_r$
\begin{center}
{\tiny
\begin{supertabular}{lll}
\hline
\hline
 $F_{i-pqr}[1]=\sqrt{\gamma_p}D_{i-qsr}[1]\e^{-\gamma_pT_p}\bigl[Z+\gamma_p+d_{i-qsr}[1]\bigr]^{-1}$  
&$f_{i-pqr}[1]=Z+\gamma_p+d_{i-qsr}[1]$ 
&$\tau_{i-pqr}[1]=T_r$                            \\
 $F_{m-pqr}[1]=\sqrt{\gamma_p}D_{m-qsr}[1]\e^{-\gamma_pT_p}\bigl[Z+\gamma_p+d_{m-qsr}[1]\bigr]^{-1}$ 
&$f_{m-pqr}[1]=Z+\gamma_p+d_{m-qsr}[1]$   
&$\tau_{m-pqr-1}[1]=T_r\quad\tau_{m-pqr-2}[1]=T_q$                             \\
 $F_{m-pqr}[2]=\sqrt{\gamma_p}D_{m-qsr}[2]\e^{-\gamma_pT_p}\bigl[Z+\gamma_p+d_{m-qsr}[2]\bigr]^{-1}$ 
&$f_{m-pqr}[2]=Z+\gamma_p+d_{m-qsr}[2]$   
&$\tau_{m-pqr-1}[2]=T_r\quad\tau_{m-pqr-2}[2]=T_q$                             \\
 $F_{m-pqr}[3]=\sqrt{\gamma_p}D_{s-qsr}[1]\e^{-\gamma_pT_p}\bigl[Z+\gamma_p+d_{s-qsr}[1]\bigr]^{-1}$
&$f_{m-pqr}[3]=Z+\gamma_p+d_{s-qsr}[1]$
&$\tau_{m-pqr-1}[3]=T_r\quad\tau_{m-pqr-2}[3]=T_p$        \\
 $F_{m-pqr}[4]=\sqrt{\gamma_p}D_{s-qsr}[2]\e^{-\gamma_pT_p}\bigl[Z+\gamma_p+d_{s-qsr}[2]\bigr]^{-1}$
&$f_{m-pqr}[4]=Z+\gamma_p+d_{s-qsr}[2]$
&$\tau_{m-pqr-1}[4]=T_q\quad\tau_{m-pqr-2}[4]=T_p$        \\
 $F_{m-pqr}[5]=\sqrt{\gamma_p}D_{s-qsr}[3]\e^{-\gamma_pT_p}\bigl[Z+\gamma_p+d_{s-qsr}[3]\bigr]^{-1}$
&$f_{m-pqr}[5]=Z+\gamma_p+d_{s-qsr}[3]$
&$\tau_{m-pqr-1}[5]=T_q\quad\tau_{m-pqr-2}[5]=T_p$        \\
 $F_{m-pqr}[6]=\sqrt{\gamma_p}D_{s-qsr}[4]\e^{-\gamma_pT_p}\bigl[Z+\gamma_p+d_{s-qsr}[4]\bigr]^{-1}$
&$f_{m-pqr}[6]=Z+\gamma_p+d_{s-qsr}[4]$
&$\tau_{m-pqr-1}[6]=T_q\quad\tau_{m-pqr-2}[6]=T_p$        \\
 $F_{s-pqr}[1]=\sqrt{\gamma_p}\Bigl[D_{i-qsr}[1]\e^{(Z+\gamma_p+d_{i-qsr}[1])T_r-\gamma_pT_p}\bigl[Z+\gamma_p+d_{i-qsr}[1]\bigr]^{-1}$
&$f_{s-pqr}[1]=0$
&$\tau_{s-pqr}[1]=T_r$        \\
 $\qquad-\sum_{\alpha=1}^2 D_{m-qsr}[\alpha]\e^{(Z+\gamma_p+d_{m-qsr}[\alpha])T_r-\gamma_pT_p}\bigl[Z+\gamma_p+d_{m-qsr}[\alpha]\bigr]^{-1}$
&$ $
&$ $        \\
 $\qquad-D_{s-qsr}[1]\e^{(Z+\gamma_p+d_{s-qsr}[1])T_r-\gamma_pT_p}\bigl[Z+\gamma_p+d_{s-qsr}[1]\bigr]^{-1}\Bigr]$
&$ $
&$ $        \\
 $F_{s-pqr}[2]=\sqrt{\gamma_p}\Bigl[\sum_{\alpha=1}^2 D_{m-qsr}[\alpha]\e^{(Z+\gamma_p+d_{m-qsr}[\alpha])T_q-\gamma_pT_p}\bigl[Z+\gamma_p+d_{m-qsr}[\alpha]\bigr]^{-1}$
&$f_{s-pqr}[2]=0$
&$\tau_{s-pqr}[2]=T_q$        \\
 $\qquad+\sum_{\alpha=2}^4 D_{s-qsr}[\alpha]\e^{(Z+\gamma_p+d_{s-qsr}[\alpha])T_r-\gamma_pT_p}\bigl[Z+\gamma_p+d_{s-qsr}[\alpha]\bigr]^{-1}\Bigr]$
&$ $
&$ $        \\
 $F_{s-pqr}[3]=\sqrt{\gamma_p}\Bigl[-\sum_{\alpha=1}^4 D_{s-qsr}[\alpha]\e^{(Z+d_{s-qsr}[\alpha])T_p}\bigl[Z-\gamma_p+d_{s-qsr}[\alpha]\bigr]^{-1}$
&$f_{s-pqr}[3]=0$
&$\tau_{s-pqr}[3]=T_p$        \\
 $\qquad+\sum_{\alpha=1}^4 D_{s-qsr}[\alpha]\e^{(Z+d_{s-qsr}[\alpha])T_p}\bigl[Z+\gamma_p+d_{s-qsr}[\alpha]\bigr]^{-1}    \Bigr]$
&$ $
&$ $        \\
 $F_{s-pqr}[4]=\sqrt{\gamma_p}D_{s-qsr}[1]\e^{\gamma_pT_p}\bigl[Z-\gamma_p+d_{s-qsr}[1]\bigr]^{-1}$
&$f_{s-pqr}[4]=Z-\gamma_p+d_{s-qsr}[1]$
&$\tau_{s-pqr}[4]=T_p$        \\
 $F_{s-pqr}[5]=\sqrt{\gamma_p}D_{s-qsr}[2]\e^{\gamma_pT_p}\bigl[Z-\gamma_p+d_{s-qsr}[2]\bigr]^{-1}$
&$f_{s-pqr}[5]=Z-\gamma_p+d_{s-qsr}[2]$
&$\tau_{s-pqr}[5]=T_p$        \\
 $F_{s-pqr}[6]=\sqrt{\gamma_p}D_{s-qsr}[3]\e^{\gamma_pT_p}\bigl[Z-\gamma_p+d_{s-qsr}[3]\bigr]^{-1}$
&$f_{s-pqr}[6]=Z-\gamma_p+d_{s-qsr}[3]$
&$\tau_{s-pqr}[6]=T_p$        \\
 $F_{s-pqr}[7]=\sqrt{\gamma_p}D_{s-qsr}[4]\e^{\gamma_pT_p}\bigl[Z-\gamma_p+d_{s-qsr}[4]\bigr]^{-1}$
&$f_{s-pqr}[7]=Z-\gamma_p+d_{s-qsr}[4]$
&$\tau_{s-pqr}[7]=T_p$        \\
\hline
\hline
\end{supertabular}}  
\end{center}
Second case: $T_q>T_p>T_r$
\begin{center}
{\tiny
\begin{supertabular}{lll}
\hline
\hline
 $F_{i-qpr}[1]=\sqrt{\gamma_p}D_{i-qsr}[1]\e^{-\gamma_pT_p}\bigl[Z+\gamma_p+d_{i-qsr}[1]\bigr]^{-1}$  
&$f_{i-qpr}[1]=Z+\gamma_p+d_{i-qsr}[1]$ 
&$\tau_{i-qpr}[1]=T_r$                             \\
 $F_{m-qpr}[1]=\sqrt{\gamma_p}D_{m-qsr}[1]\e^{-\gamma_pT_p}\bigl[Z+\gamma_p+d_{m-qsr}[1]\bigr]^{-1}$ 
&$f_{m-qpr}[1]=Z+\gamma_p+d_{m-qsr}[1]$   
&$\tau_{m-qpr-1}[1]=T_r\quad\tau_{m-qpr-2}[1]=T_p$                             \\
 $F_{m-qpr}[2]=\sqrt{\gamma_p}D_{m-qsr}[2]\e^{-\gamma_pT_p}\bigl[Z+\gamma_p+d_{m-qsr}[2]\bigr]^{-1}$ 
&$f_{m-qpr}[2]=Z+\gamma_p+d_{m-qsr}[2]$   
&$\tau_{m-qpr-1}[2]=T_r\quad\tau_{m-qpr-2}[2]=T_p$                             \\
 $F_{m-qpr}[3]=\sqrt{\gamma_p}D_{m-qsr}[1]\e^{\gamma_pT_p}\bigl[Z-\gamma_p+d_{m-qsr}[1]\bigr]^{-1}$ 
&$f_{m-qpr}[3]=Z-\gamma_p+d_{m-qsr}[1]$   
&$\tau_{m-qpr-1}[3]=T_p\quad\tau_{m-qpr-2}[3]=T_q$                             \\
 $F_{m-qpr}[4]=\sqrt{\gamma_p}D_{m-qsr}[2]\e^{\gamma_pT_p}\bigl[Z-\gamma_p+d_{m-qsr}[2]\bigr]^{-1}$ 
&$f_{m-qpr}[4]=Z-\gamma_p+d_{m-qsr}[2]$   
&$\tau_{m-qpr-1}[4]=T_p\quad\tau_{m-qpr-2}[4]=T_q$                             \\
 $F_{s-qpr}[1]=\sqrt{\gamma_p}\Bigl[D_{i-qsr}[1]\e^{(Z+\gamma_p+d_{i-qsr}[1])T_r-\gamma_pT_p}\bigl[Z+\gamma_p+d_{i-qsr}[1]\bigr]^{-1}$
&$f_{s-qpr}[1]=0$
&$\tau_{s-qpr}[1]=T_r$        \\
 $\qquad-\sum_{\alpha=1}^2 D_{m-qsr}[\alpha]\e^{(Z+\gamma_p+d_{m-qsr}[\alpha])T_r-\gamma_pT_p}\bigl[Z+\gamma_p+d_{m-qsr}[\alpha]\bigr]^{-1}$
&$ $
&$ $        \\
 $\qquad-D_{s-qsr}[1]\e^{(Z+\gamma_p+d_{s-qsr}[1])T_r-\gamma_pT_p}\bigl[Z+\gamma_p+d_{s-qsr}[1]\bigr]^{-1}\Bigr]$
&$ $
&$ $        \\
 $F_{s-qpr}[2]=\sqrt{\gamma_p}\Bigl[\sum_{\alpha=1}^2 D_{m-qsr}[\alpha]\e^{(Z-\gamma_p+d_{m-qsr}[\alpha])T_q+\gamma_pT_p}\bigl[Z-\gamma_p+d_{m-qsr}[\alpha]\bigr]^{-1}$
&$f_{s-qpr}[2]=0$
&$\tau_{s-qpr}[2]=T_q$        \\
 $\qquad-\sum_{\alpha=2}^4 D_{s-qsr}[\alpha]\e^{(Z-\gamma_p+d_{s-qsr}[\alpha])T_q+\gamma_pT_p}\bigl[Z-\gamma_p+d_{s-qsr}[\alpha]\bigr]^{-1}\Bigr]$
&$ $
&$ $        \\
 $F_{s-qpr}[3]=\sqrt{\gamma_p}\Bigl[-\sum_{\alpha=1}^2 D_{m-qsr}[\alpha]\e^{(Z+d_{m-qsr}[\alpha])T_p}\bigl[Z-\gamma_p+d_{m-qsr}[\alpha]\bigr]^{-1}$
&$f_{s-qpr}[3]=0$
&$\tau_{s-qpr}[3]=T_p$        \\
 $\qquad+\sum_{\alpha=1}^2 D_{m-qsr}[\alpha]\e^{(Z+d_{m-qsr}[\alpha])T_p}\bigl[Z+\gamma_p+d_{m-qsr}[\alpha]\bigr]^{-1}$
&$ $
&$ $        \\
 $\qquad+D_{s-qsr}[1]\e^{(Z+d_{s-qsr}[1])T_p}\bigl[Z+\gamma_p+d_{s-qsr}[1]\bigr]^{-1}\Bigr]$
&$ $
&$ $        \\
 $F_{s-qpr}[4]=\sqrt{\gamma_p}D_{s-qsr}[1]\e^{-\gamma_pT_p}\bigl[Z+\gamma_p+d_{s-qsr}[1]\bigr]^{-1}$
&$f_{s-qpr}[4]=Z+\gamma_p+d_{s-qsr}[1]$
&$\tau_{s-qpr}[4]=T_r$        \\
 $F_{s-qpr}[5]=\sqrt{\gamma_p}D_{s-qsr}[1]\e^{\gamma_pT_p}\bigl[Z-\gamma_p+d_{s-qsr}[1]\bigr]^{-1}$
&$f_{s-qpr}[5]=Z-\gamma_p+d_{s-qsr}[1]$
&$\tau_{s-qpr}[5]=T_p$        \\
$F_{s-qpr}[6]=\sqrt{\gamma_p}D_{s-qsr}[2]\e^{\gamma_pT_p}\bigl[Z+\gamma_p+d_{s-qsr}[2]\bigr]^{-1}$
&$f_{s-qpr}[6]=Z-\gamma_p+d_{s-qsr}[2]$
&$\tau_{s-qpr}[6]=T_q$        \\
 $F_{s-qpr}[7]=\sqrt{\gamma_p}D_{s-qsr}[3]\e^{\gamma_pT_p}\bigl[Z-\gamma_p+d_{s-qsr}[3]\bigr]^{-1}$
&$f_{s-qpr}[7]=Z-\gamma_p+d_{s-qsr}[3]$
&$\tau_{s-qpr}[7]=T_q$        \\
 $F_{s-qpr}[8]=\sqrt{\gamma_p}D_{s-qsr}[4]\e^{\gamma_pT_p}\bigl[Z-\gamma_p+d_{s-qsr}[4]\bigr]^{-1}$
&$f_{s-qpr}[8]=Z-\gamma_p+d_{s-qsr}[4]$
&$\tau_{s-qpr}[8]=T_q$        \\
 $F_{s-qpr}[9]=-\sqrt{\gamma_p}D_{s-qsr}[1]\e^{-\gamma_pT_p}\bigl[Z+\gamma_p+d_{s-qsr}[1]\bigr]^{-1}$
&$f_{s-qpr}[9]=Z+\gamma_p+d_{s-qsr}[1]$
&$\tau_{s-qpr}[9]=T_p$        \\
\hline
\hline
\end{supertabular}}  
\end{center}
\end{minipage}}\newpage
\newpage\rotatebox{90}{\begin{minipage}{\textheight}
Third case: $T_p>T_r>T_q$
\begin{center}
{\tiny
\begin{supertabular}{lll}
\hline
\hline
 $F_{i-prq}[1]=\sqrt{\gamma_p}D_{i-rsq}[1]\e^{-\gamma_pT_p}\bigl[Z+\gamma_p+d_{i-rsq}[1]\bigr]^{-1}$  
&$f_{i-prq}[1]=Z+\gamma_p+d_{i-rsq}[1]$ 
&$\tau_{i-prq}[1]=T_q$                             \\
 $F_{m-prq}[1]=\sqrt{\gamma_p}D_{m-rsq}[1]\e^{-\gamma_pT_p}\bigl[Z+\gamma_p+d_{m-rsq}[1]\bigr]^{-1}$ 
&$f_{m-prq}[1]=Z+\gamma_p+d_{m-rsq}[1]$   
&$\tau_{m-prq-1}[1]=T_q\quad\tau_{m-prq-2}[1]=T_r$                             \\
 $F_{m-prq}[2]=\sqrt{\gamma_p}D_{s-rsq}[1]\e^{-\gamma_pT_p}\bigl[Z+\gamma_p+d_{s-rsq}[1]\bigr]^{-1}$
&$f_{m-prq}[2]=Z+\gamma_p+d_{s-rsq}[1]$
&$\tau_{m-prq-1}[2]=T_q\quad\tau_{m-prq-2}[2]=T_p$        \\
 $F_{m-prq}[3]=\sqrt{\gamma_p}D_{s-rsq}[2]\e^{-\gamma_pT_p}\bigl[Z+\gamma_p+d_{s-rsq}[2]\bigr]^{-1}$
&$f_{m-prq}[3]=Z+\gamma_p+d_{s-rsq}[2]$
&$\tau_{m-prq-1}[3]=T_r\quad\tau_{m-prq-2}[3]=T_p$        \\
 $F_{m-prq}[4]=\sqrt{\gamma_p}D_{s-rsq}[3]\e^{-\gamma_pT_p}\bigl[Z+\gamma_p+d_{s-rsq}[3]\bigr]^{-1}$
&$f_{m-prq}[4]=Z+\gamma_p+d_{s-rsq}[3]$
&$\tau_{m-prq-1}[4]=T_r\quad\tau_{m-prq-2}[4]=T_p$        \\
 $F_{m-prq}[5]=\sqrt{\gamma_p}D_{s-rsq}[4]\e^{-\gamma_pT_p}\bigl[Z+\gamma_p+d_{s-rsq}[4]\bigr]^{-1}$
&$f_{m-prq}[5]=Z+\gamma_p+d_{s-rsq}[4]$
&$\tau_{m-prq-1}[5]=T_r\quad\tau_{m-prq-2}[5]=T_p$        \\
 $F_{s-prq}[1]=\sqrt{\gamma_p}\Bigl[D_{m-rsq}[1]\e^{(Z+\gamma_p+d_{m-rsq}[1])T_r-\gamma_pT_p}\bigl[Z+\gamma_p+d_{m-rsq}[1]\bigr]^{-1}$
&$f_{s-prq}[1]=0$
&$\tau_{s-prq}[1]=T_r$        \\
 $\qquad-\sum_{\alpha=2}^4 D_{s-rsq}[\alpha]\e^{(Z+\gamma_p+d_{s-rsq}[\alpha])T_r-\gamma_pT_p}\bigl[Z+\gamma_p+d_{s-rsq}[\alpha]\bigr]^{-1}\Bigr]$
&$ $
&$ $        \\
 $F_{s-prq}[2]=\sqrt{\gamma_p}\Bigl[D_{i-rsq}[1]\e^{(Z+\gamma_p+d_{i-rsq}[1])T_q-\gamma_pT_p}\bigl[Z+\gamma_p+d_{i-rsq}[1]\bigr]^{-1}$
&$f_{s-prq}[2]=0$
&$\tau_{s-prq}[2]=T_q$        \\
 $\qquad +D_{m-rsq}[1]\e^{(Z+\gamma_p+d_{m-rsq}[1])T_q-\gamma_pT_p}\bigl[Z+\gamma_p+d_{m-rsq}[1]\bigr]^{-1}$
&$ $
&$ $        \\
 $\qquad-D_{s-rsq}[1]\e^{(Z+\gamma_p+d_{s-rsq}[1])T_q-\gamma_pT_p}\bigl[Z+\gamma_p+d_{s-rsq}[1]\bigr]^{-1}\Bigr]$
&$ $
&$ $        \\
 $F_{s-prq}[3]=\sqrt{\gamma_p}\Bigl[\sum_{\alpha=1}^4 D_{s-rsq}[\alpha]\e^{(Z+d_{s-rsq}[\alpha])T_p}\bigl[Z+\gamma_p+d_{s-rsq}[\alpha]\bigr]^{-1}$
&$f_{s-prq}[3]=0$
&$\tau_{s-prq}[3]=T_p$        \\
 $\qquad-\sum_{\alpha=1}^4 D_{s-rsq}[\alpha]\e^{(Z+d_{s-rsq}[\alpha])T_p}\bigl[Z-\gamma_p+d_{s-rsq}[\alpha]\bigr]^{-1}$
&$ $
&$ $        \\
 $F_{s-prq}[4]=\sqrt{\gamma_p}D_{s-rsq}[1]\e^{\gamma_pT_p}\bigl[Z-\gamma_p+d_{s-rsq}[1]\bigr]^{-1}$
&$f_{s-prq}[4]=Z-\gamma_p+d_{s-rsq}[1]$
&$\tau_{s-prq}[4]=T_p$        \\
 $F_{s-prq}[5]=\sqrt{\gamma_p}D_{s-rsq}[2]\e^{\gamma_pT_p}\bigl[Z-\gamma_p+d_{s-rsq}[2]\bigr]^{-1}$
&$f_{s-prq}[5]=Z-\gamma_p+d_{s-rsq}[2]$
&$\tau_{s-prq}[5]=T_p$        \\
 $F_{s-prq}[6]=\sqrt{\gamma_p}D_{s-rsq}[3]\e^{\gamma_pT_p}\bigl[Z-\gamma_p+d_{s-rsq}[3]\bigr]^{-1}$
&$f_{s-prq}[6]=Z-\gamma_p+d_{s-rsq}[3]$
&$\tau_{s-prq}[6]=T_p$        \\
 $F_{s-prq}[7]=\sqrt{\gamma_p}D_{s-rsq}[4]\e^{\gamma_pT_p}\bigl[Z-\gamma_p+d_{s-rsq}[4]\bigr]^{-1}$
&$f_{s-prq}[7]=Z-\gamma_p+d_{s-rsq}[4]$
&$\tau_{s-prq}[7]=T_p$          \\
\hline
\hline
\end{supertabular}}  
\end{center}
Fourth case: $T_r>T_p>T_q$
\begin{center}
{\tiny
\begin{supertabular}{lll}
\hline
\hline
 $F_{i-rpq}[1]=\sqrt{\gamma_p}D_{i-rsq}[1]\e^{-\gamma_pT_p}\bigl[Z+\gamma_p+d_{i-rsq}[1]\bigr]^{-1}$  
&$f_{i-rpq}[1]=Z+\gamma_p+d_{i-rsq}[1]$ 
&$\tau_{i-rpq}[1]=T_q$                             \\
 $F_{m-rpq}[1]=\sqrt{\gamma_p}D_{m-rsq}[1]\e^{-\gamma_pT_p}\bigl[Z+\gamma_p+d_{m-rsq}[1]\bigr]^{-1}$ 
&$f_{m-rpq}[1]=Z+\gamma_p+d_{m-rsq}[1]$   
&$\tau_{m-rpq-1}[1]=T_q\quad\tau_{m-rpq-2}[1]=T_p$                             \\
 $F_{m-rpq}[2]=\sqrt{\gamma_p}D_{m-rsq}[1]\e^{\gamma_pT_p}\bigl[Z-\gamma_p+d_{m-rsq}[1]\bigr]^{-1}$ 
&$f_{m-rpq}[2]=Z-\gamma_p+d_{m-rsq}[1]$   
&$\tau_{m-rpq-1}[2]=T_p\quad\tau_{m-rpq-2}[1]=T_r$                             \\
 $F_{m-rpq}[3]=\sqrt{\gamma_p}D_{s-rsq}[1]\e^{-\gamma_pT_p}\bigl[Z+\gamma_p+d_{s-rsq}[1]\bigr]^{-1}$ 
&$f_{m-rpq}[3]=Z+\gamma_p+d_{s-rsq}[1]$   
&$\tau_{m-rpq-1}[3]=T_q\quad\tau_{m-rpq-2}[1]=T_p$                             \\
 $F_{s-rpq}[1]=\sqrt{\gamma_p}\Bigl[D_{m-rsq}[1]\e^{(Z-\gamma_p+d_{m-rsq}[1])T_r+\gamma_pT_p}\bigl[Z-\gamma_p+d_{m-rsq}[1]\bigr]^{-1}$
&$f_{s-rpq}[1]=0$
&$\tau_{s-rpq}[1]=T_r$                                                           \\
 $\qquad-\sum_{\alpha=2}^4 D_{s-rsq}[\alpha]\e^{(Z-\gamma_p+d_{s-rsq}[\alpha])T_r+\gamma_pT_p}\bigl[Z-\gamma_p+d_{s-rsq}[\alpha]\bigr]^{-1}\Bigr]$
&$ $
&$ $                                                                           \\
 $F_{s-rpq}[2]=\sqrt{\gamma_p}\Bigl[D_{i-rsq}[1]\e^{(Z+\gamma_p+d_{i-rsq}[1])T_q-\gamma_pT_p}\bigl[Z+\gamma_p+d_{i-rsq}[1]\bigr]^{-1}$
&$f_{s-rpq}[2]=0$
&$\tau_{s-rpq}[2]=T_q$                                                        \\
 $\qquad -D_{m-rsq}[1]\e^{(Z+\gamma_p+d_{m-rsq}[1])T_q-\gamma_pT_p}\bigl[Z+\gamma_p+d_{m-rsq}[1]\bigr]^{-1}$
&$ $
&$ $                                                                             \\
 $\qquad-D_{s-rsq}[1]\e^{(Z+\gamma_p+d_{s-rsq}[1])T_q-\gamma_pT_p}\bigl[Z+\gamma_p+d_{s-rsq}[1]\bigr]^{-1}\Bigr]$
&$ $
&$ $                                                                            \\
 $F_{s-rpq}[3]=\sqrt{\gamma_p}\Bigl[D_{m-rsq}[1]\e^{(Z+d_{m-rsq}[1])T_p}\bigl[Z+\gamma_p+d_{m-rsq}[1]\bigr]^{-1}$
&$f_{s-rpq}[3]=0$
&$\tau_{s-rpq}[3]=T_p$        \\
 $\qquad-D_{m-rsq}[1]\e^{(Z+d_{m-rsq}[1])T_p}\bigl[Z-\gamma_p+d_{m-rsq}[1]\bigr]^{-1}$
&$ $
&$ $                                                                              \\
 $\qquad D_{s-rsq}[1]\e^{(Z+d_{s-rsq}[1])T_p}\bigl[Z+\gamma_p+d_{s-rsq}[1]\bigr]^{-1}$
&$ $
&$ $                                                                             \\
 $\qquad-D_{s-rsq}[1]\e^{(Z+d_{s-rsq}[1])T_p}\bigl[Z-\gamma_p+d_{s-rsq}[1]\bigr]^{-1}$
&$ $
&$ $                                                                               \\
 $F_{s-rpq}[4]=\sqrt{\gamma_p}D_{s-rsq}[1]\e^{\gamma_pT_p}\bigl[Z-\gamma_p+d_{s-rsq}[1]\bigr]^{-1}$
&$f_{s-rpq}[4]=Z-\gamma_p+d_{s-rsq}[1]$
&$\tau_{s-rpq}[4]=T_p$                                                              \\
 $F_{s-rpq}[5]=\sqrt{\gamma_p}D_{s-rsq}[2]\e^{\gamma_pT_p}\bigl[Z-\gamma_p+d_{s-rsq}[2]\bigr]^{-1}$
&$f_{s-rpq}[5]=Z-\gamma_p+d_{s-rsq}[2]$
&$\tau_{s-rpq}[5]=T_r$                                                              \\
 $F_{s-rpq}[6]=\sqrt{\gamma_p}D_{s-rsq}[3]\e^{\gamma_pT_p}\bigl[Z-\gamma_p+d_{s-rsq}[3]\bigr]^{-1}$
&$f_{s-rpq}[6]=Z-\gamma_p+d_{s-rsq}[3]$
&$\tau_{s-rpq}[6]=T_r$                                                              \\
 $F_{s-rpq}[7]=\sqrt{\gamma_p}D_{s-rsq}[4]\e^{\gamma_pT_p}\bigl[Z-\gamma_p+d_{s-rsq}[4]\bigr]^{-1}$
&$f_{s-rpq}[7]=Z-\gamma_p+d_{s-rsq}[4]$
&$\tau_{s-rpq}[7]=T_r$                                                              \\
\hline
\hline
\end{supertabular}}  
\end{center}
\end{minipage}}\newpage
\newpage\rotatebox{90}{\begin{minipage}{\textheight}
Fifth case: $T_q>T_r>T_p$
\begin{center}
{\tiny
\begin{supertabular}{lll}
\hline
\hline
 $F_{i-qrp}[1]=\sqrt{\gamma_p}D_{i-qsr}[1]\e^{-\gamma_pT_p}\bigl[Z+\gamma_p+d_{i-qsr}[1]\bigr]^{-1}$  
&$f_{i-qrp}[1]=Z+\gamma_p+d_{i-qsr}[1]$ 
&$\tau_{i-qrp}[1]=T_p$                             \\
 $F_{m-qrp}[1]=\sqrt{\gamma_p}D_{m-qsr}[1]\e^{\gamma_pT_p}\bigl[Z-\gamma_p+d_{m-qsr}[1]\bigr]^{-1}$ 
&$f_{m-qrp}[1]=Z-\gamma_p+d_{m-qsr}[1]$   
&$\tau_{m-qrp-1}[1]=T_r\quad\tau_{m-qrp-2}[1]=T_q$                             \\
 $F_{m-qrp}[2]=\sqrt{\gamma_p}D_{m-qsr}[2]\e^{\gamma_pT_p}\bigl[Z-\gamma_p+d_{m-qsr}[2]\bigr]^{-1}$ 
&$f_{m-qrp}[2]=Z-\gamma_p+d_{m-qsr}[2]$   
&$\tau_{m-qrp-1}[2]=T_r\quad\tau_{m-qrp-2}[1]=T_q$                             \\
 $F_{m-qrp}[3]=\sqrt{\gamma_p}D_{i-qsr}[1]\e^{\gamma_pT_p}\bigl[Z-\gamma_p+d_{i-qsr}[1]\bigr]^{-1}$
&$f_{m-qrp}[3]=Z-\gamma_p+d_{i-qsr}[1]$
&$\tau_{m-qrp-1}[3]=T_p\quad\tau_{m-qrp-2}[4]=T_r$                           \\
 $F_{s-qrp}[1]=\sqrt{\gamma_p}\Bigl[D_{i-qsr}[1]\e^{(Z-\gamma_p+d_{i-qsr}[1])T_r+\gamma_pT_p}\bigl[Z-\gamma_p+d_{i-qsr}[1]\bigr]^{-1}$
&$f_{s-qrp}[1]=0$
&$\tau_{s-qrp}[1]=T_r$                                                        \\
 $\qquad-D_{s-qsr}[1]\e^{(Z-\gamma_p+d_{s-qsr}[1])T_r+\gamma_pT_p}\bigl[Z-\gamma_p+d_{s-qsr}[1]\bigr]^{-1}\Bigr]$
&$ $
&$ $                                                                           \\
 $\qquad-\sum_{\alpha=1}^2 D_{m-qsr}[\alpha]\e^{(Z-\gamma_p+d_{m-qsr}[\alpha])T_r+\gamma_pT_p}\bigl[Z-\gamma_p+d_{m-qsr}[\alpha]\bigr]^{-1}\Bigr]$
&$ $
&$ $                                                                           \\
 $F_{s-qrp}[2]=\sqrt{\gamma_p}\Bigl[\sum_{\alpha=1}^2 D_{m-qsr}[\alpha]\e^{(Z-\gamma_p+d_{m-qsr}[\alpha])T_q+\gamma_pT_p}\bigl[Z-\gamma_p+d_{m-qsr}[\alpha]\bigr]^{-1}$     
&$f_{s-qrp}[2]=0$
&$\tau_{s-qrp}[2]=T_q$                                                        \\
 $\qquad -\sum_{\alpha=2}^4 D_{s-qsr}[\alpha]\e^{(Z-\gamma_p+d_{s-qsr}[\alpha])T_q+\gamma_pT_p}\bigl[Z-\gamma_p+d_{s-qsr}[\alpha]\bigr]^{-1}$
&$ $
&$ $                                                                             \\
 $F_{s-qrp}[3]=\sqrt{\gamma_p}\Bigl[D_{i-qsr}[1]\e^{(Z+d_{i-qsr}[1])T_p}\bigl[Z+\gamma_p+d_{i-qsr}[1]\bigr]^{-1}$
&$f_{s-qrp}[3]=0$
&$\tau_{s-qrp}[3]=T_p$        \\
 $\qquad -D_{i-qsr}[1]\e^{(Z+d_{i-qsr}[1])T_p}\bigl[Z-\gamma_p+d_{i-qsr}[1]\bigr]^{-1}$
&$ $
&$ $                                                                              \\
 $F_{s-qrp}[4]=\sqrt{\gamma_p}D_{s-qsr}[1]\e^{\gamma_pT_p}\bigl[Z-\gamma_p+d_{s-qsr}[1]\bigr]^{-1}$
&$f_{s-qrp}[4]=Z-\gamma_p+d_{s-qsr}[1]$
&$\tau_{s-qrp}[4]=T_r$                                                              \\
 $F_{s-qrp}[5]=\sqrt{\gamma_p}D_{s-qsr}[2]\e^{\gamma_pT_p}\bigl[Z-\gamma_p+d_{s-qsr}[2]\bigr]^{-1}$
&$f_{s-qrp}[5]=Z-\gamma_p+d_{s-qsr}[2]$
&$\tau_{s-qrp}[5]=T_q$                                                              \\
 $F_{s-qrp}[6]=\sqrt{\gamma_p}D_{s-qsr}[3]\e^{\gamma_pT_p}\bigl[Z-\gamma_p+d_{s-qsr}[3]\bigr]^{-1}$
&$f_{s-qrp}[6]=Z-\gamma_p+d_{s-qsr}[3]$
&$\tau_{s-qrp}[6]=T_q$                                                              \\
 $F_{s-qrp}[7]=\sqrt{\gamma_p}D_{s-qsr}[4]\e^{\gamma_pT_p}\bigl[Z-\gamma_p+d_{s-qsr}[4]\bigr]^{-1}$
&$f_{s-qrp}[7]=Z-\gamma_p+d_{s-qsr}[4]$
&$\tau_{s-qrp}[7]=T_q$                                                              \\
\hline
\hline
\end{supertabular}}  
\end{center}
Sixth case: $T_r>T_p>T_q$
\begin{center}
\tiny{
\begin{supertabular}{lll}
\hline
\hline
 $F_{i-rqp}[1]=\sqrt{\gamma_p}D_{i-rsq}[1]\e^{-\gamma_pT_p}\bigl[Z+\gamma_p+d_{i-rsq}[1]\bigr]^{-1}$  
&$f_{i-rqp}[1]=Z+\gamma_p+d_{i-rsq}[1]$ 
&$\tau_{i-rqp}[1]=T_p$                             \\
 $F_{m-rqp}[1]=\sqrt{\gamma_p}D_{m-rsq}[1]\e^{\gamma_pT_p}\bigl[Z-\gamma_p+d_{m-rsq}[1]\bigr]^{-1}$ 
&$f_{m-rqp}[1]=Z-\gamma_p+d_{m-rsq}[1]$   
&$\tau_{m-rqp-1}[1]=T_q\quad\tau_{m-rqp-2}[1]=T_r$                             \\
 $F_{m-rqp}[2]=\sqrt{\gamma_p}D_{i-rsq}[1]\e^{\gamma_pT_p}\bigl[Z-\gamma_p+d_{i-rsq}[1]\bigr]^{-1}$ 
&$f_{m-rqp}[2]=Z-\gamma_p+d_{i-rsq}[1]$   
&$\tau_{m-rqp-1}[2]=T_p\quad\tau_{m-rqp-2}[1]=T_q$                             \\
 $F_{s-rqp}[1]=\sqrt{\gamma_p}\Bigl[D_{m-rsq}[1]\e^{(Z-\gamma_p+d_{m-rsq}[1])T_r+\gamma_pT_p}\bigl[Z-\gamma_p+d_{m-rsq}[1]\bigr]^{-1}$
&$f_{s-rqp}[1]=0$
&$\tau_{s-rqp}[1]=T_r$                                                        \\
 $\qquad-\sum_{\alpha=2}^4 D_{s-rsq}[\alpha]\e^{(Z-\gamma_p+d_{s-rsq}[\alpha])T_r+\gamma_pT_p}\bigl[Z-\gamma_p+d_{s-rsq}[\alpha]\bigr]^{-1}\Bigr]$
&$ $
&$ $                                                                           \\
 $F_{s-rqp}[2]=\sqrt{\gamma_p}\Bigl[ D_{i-rsq}[1]\e^{(Z-\gamma_p+d_{i-rsq}[1])T_q+\gamma_pT_p}\bigl[Z-\gamma_p+d_{i-rsq}[1]\bigr]^{-1}$     
&$f_{s-rqp}[2]=0$
&$\tau_{s-rqp}[2]=T_q$                                                        \\
 $\qquad -D_{m-rsq}[1]\e^{(Z-\gamma_p+d_{m-rsq}[1])T_q+\gamma_pT_p}\bigl[Z-\gamma_p+d_{m-rsq}[1]\bigr]^{-1}$
&$ $
&$ $                                                                             \\
 $\qquad -D_{s-rsq}[1]\e^{(Z-\gamma_p+d_{s-rsq}[1])T_q+\gamma_pT_p}\bigl[Z-\gamma_p+d_{s-rsq}[1]\bigr]^{-1} \Bigr]$
&$ $
&$ $                                                                             \\
 $F_{s-rqp}[3]=\sqrt{\gamma_p}\Bigl[D_{i-rsq}[1]\e^{(Z+d_{i-rsq}[1])T_p}\bigl[Z+\gamma_p+d_{i-rsq}[1]\bigr]^{-1}$
&$f_{s-rqp}[3]=0$
&$\tau_{s-rqp}[3]=T_p$        \\
 $\qquad -D_{i-rsq}[1]\e^{(Z+d_{i-rsq}[1])T_p}\bigl[Z-\gamma_p+d_{i-rsq}[1]\bigr]^{-1}$
&$ $
&$ $                                                                              \\
 $F_{s-rqp}[4]=\sqrt{\gamma_p}D_{s-rsq}[1]\e^{\gamma_pT_p}\bigl[Z-\gamma_p+d_{s-rsq}[1]\bigr]^{-1}$
&$f_{s-rqp}[4]=Z-\gamma_p+d_{s-rsq}[1]$
&$\tau_{s-rqp}[4]=T_q$                                                              \\
 $F_{s-rqp}[5]=\sqrt{\gamma_p}D_{s-rsq}[2]\e^{\gamma_pT_p}\bigl[Z-\gamma_p+d_{s-rsq}[2]\bigr]^{-1}$
&$f_{s-rqp}[5]=Z-\gamma_p+d_{s-rsq}[2]$
&$\tau_{s-rqp}[5]=T_r$                                                              \\
 $F_{s-rqp}[6]=\sqrt{\gamma_p}D_{s-rsq}[3]\e^{\gamma_pT_p}\bigl[Z-\gamma_p+d_{s-rsq}[3]\bigr]^{-1}$
&$f_{s-rqp}[6]=Z-\gamma_p+d_{s-rsq}[3]$
&$\tau_{s-rqp}[6]=T_r$                                                              \\
 $F_{s-rqp}[7]=\sqrt{\gamma_p}D_{s-rsq}[4]\e^{\gamma_pT_p}\bigl[Z-\gamma_p+d_{s-rsq}[4]\bigr]^{-1}$
&$f_{s-rqp}[7]=Z-\gamma_p+d_{s-rsq}[4]$
&$\tau_{s-rqp}[7]=T_r$                                                              \\
\hline  
\hline
\end{supertabular}}  
\end{center}  
\end{minipage}}\newpage

\section{Pathways involved}

We describe all the pathways associated with processes participating in the third-order polarization in the phase-matched direction  $-\vec{\g{k}}_1+\vec{\g{k}}_2+\vec{\g{k}}_3$ corresponding to photon-echo process. All these pathways are described in the eigenstates basis set $\{\vert \ep i\rangle\}$.
Some pathways whose contributions are extremely small have been omitted as indicated in the caption.
\begin{center}
{\scriptsize
\bottomcaption{Pathways participating in the 2D vibrational spectrum. Here, processes associated with dipolar couplings between states $\{\vert\ep i\rangle\leftrightarrow\vert \ep j\rangle\}$ for $(i,j)=(2,3)$, $(4,5)$, $(5,6)$ and $(4,6)$ are rejected because the magnitude of their dipole moments are order of magnitude smaller than the other ones. Also, the symbols $(\pm)$ indicate the sign of the $\vec{k}$-component of the fields which needs to be considered for the laser field component, when the rotating wave approximation (RWA) is introduced.}
\begin{supertabular}{cccccccc}
\hline
\hline
$\!\scr{\g\rho(t)}\!$&$\!\scr{\g{G}(t-\tau_3)}\!$&$\!\scr{\g{L}_{v[p]}(\tau_3)}\!$&$\!\scr{\g{G}(\tau_3-\tau_2)}\!$&$\!\scr{\g{L}_{v[q]}(\tau_2)}\!$&$\!\scr{\g{G}((\tau_2-\tau_1))}\!$ &$\!\scr{\g{L}_{v[r]}(\tau_1)}\!$&$\!\scr{\g{\rho}(t_0)}\!$\\
\hline
\hline
$\scr{\ep2\ep1}$&$\scr{\ep2\ep1\ep2\ep1}$&$\scr{\ep2\ep1\ep1\ep1}^{(+)}$&$\scr{\ep1\ep1\ep1\ep1}$&$\scr{\ep1\ep1\ep1\ep2}^{(+)}$&$\scr{\ep1\ep2\ep1\ep2}$&$\scr{\ep1\ep2\ep1\ep1}^{(-)}$&$\scr{\ep1\ep1}$\\
\hline
$\scr{\ep3\ep1}$&$\scr{\ep3\ep1\ep3\ep1}$&$\scr{\ep3\ep1\ep1\ep1}^{(+)}$&$\scr{\ep1\ep1\ep1\ep1}$&$\scr{\ep1\ep1\ep1\ep2}^{(+)}$&$\scr{\ep1\ep2\ep1\ep2}$&$\scr{\ep1\ep2\ep1\ep1}^{(-)}$&$\scr{\ep1\ep1}$\\
\hline
$\scr{\ep2\ep1}$&$\scr{\ep2\ep1\ep2\ep1}$&$\scr{\ep2\ep1\ep2\ep2}^{(+)}$&$\scr{\ep2\ep2\ep2\ep2}$&$\scr{\ep2\ep2\ep1\ep2}^{(+)}$&$\scr{\ep1\ep2\ep1\ep2}$&$\scr{\ep1\ep2\ep1\ep1}^{(-)}$&$\scr{\ep1\ep1}$\\
\hline
$\scr{\ep4\ep2}$&$\scr{\ep4\ep2\ep4\ep2}$&$\scr{\ep4\ep2\ep2\ep2}^{(+)}$&$\scr{\ep2\ep2\ep2\ep2}$&$\scr{\ep2\ep2\ep1\ep2}^{(+)}$&$\scr{\ep1\ep2\ep1\ep2}$&$\scr{\ep1\ep2\ep1\ep1}^{(-)}$&$\scr{\ep1\ep1}$\\
\hline
$\scr{\ep5\ep2}$&$\scr{\ep5\ep2\ep5\ep2}$&$\scr{\ep5\ep2\ep2\ep2}^{(+)}$&$\scr{\ep2\ep2\ep2\ep2}$&$\scr{\ep2\ep2\ep1\ep2}^{(+)}$&$\scr{\ep1\ep2\ep1\ep2}$&$\scr{\ep1\ep2\ep1\ep1}^{(-)}$&$\scr{\ep1\ep1}$\\
\hline
$\scr{\ep6\ep2}$&$\scr{\ep6\ep2\ep6\ep2}$&$\scr{\ep6\ep2\ep2\ep2}^{(+)}$&$\scr{\ep2\ep2\ep2\ep2}$&$\scr{\ep2\ep2\ep1\ep2}^{(+)}$&$\scr{\ep1\ep2\ep1\ep2}$&$\scr{\ep1\ep2\ep1\ep1}^{(-)}$&$\scr{\ep1\ep1}$\\
\hline
$\scr{\ep3\ep1}$&$\scr{\ep3\ep1\ep3\ep1}$&$\scr{\ep3\ep1\ep3\ep2}^{(+)}$&$\scr{\ep3\ep2\ep3\ep2}$&$\scr{\ep3\ep2\ep1\ep2}^{(+)}$&$\scr{\ep1\ep2\ep1\ep2}$&$\scr{\ep1\ep2\ep1\ep1}^{(-)}$&$\scr{\ep1\ep1}$\\
\hline
$\scr{\ep4\ep2}$&$\scr{\ep4\ep2\ep4\ep2}$&$\scr{\ep4\ep2\ep3\ep2}^{(+)}$&$\scr{\ep3\ep2\ep3\ep2}$&$\scr{\ep3\ep2\ep1\ep2}^{(+)}$&$\scr{\ep1\ep2\ep1\ep2}$&$\scr{\ep1\ep2\ep1\ep1}^{(-)}$&$\scr{\ep1\ep1}$\\
\hline
$\scr{\ep5\ep2}$&$\scr{\ep5\ep2\ep5\ep2}$&$\scr{\ep5\ep2\ep3\ep2}^{(+)}$&$\scr{\ep3\ep2\ep3\ep2}$&$\scr{\ep3\ep2\ep1\ep2}^{(+)}$&$\scr{\ep1\ep2\ep1\ep2}$&$\scr{\ep1\ep2\ep1\ep1}^{(-)}$&$\scr{\ep1\ep1}$\\
\hline
$\scr{\ep6\ep2}$&$\scr{\ep6\ep2\ep6\ep2}$&$\scr{\ep6\ep2\ep3\ep2}^{(+)}$&$\scr{\ep3\ep2\ep3\ep2}$&$\scr{\ep3\ep2\ep1\ep2}^{(+)}$&$\scr{\ep1\ep2\ep1\ep2}$&$\scr{\ep1\ep2\ep1\ep1}^{(-)}$&$\scr{\ep1\ep1}$\\
\hline
$\scr{\ep2\ep1}$&$\scr{\ep2\ep1\ep2\ep1}$&$\scr{\ep2\ep1\ep1\ep1}^{(+)}$&$\scr{\ep1\ep1\ep1\ep1}$&$\scr{\ep1\ep1\ep1\ep3}^{(+)}$&$\scr{\ep1\ep3\ep1\ep3}$&$\scr{\ep1\ep3\ep1\ep1}^{(-)}$&$\scr{\ep1\ep1}$\\
\hline
$\scr{\ep3\ep1}$&$\scr{\ep3\ep1\ep3\ep1}$&$\scr{\ep3\ep1\ep1\ep1}^{(+)}$&$\scr{\ep1\ep1\ep1\ep1}$&$\scr{\ep1\ep1\ep1\ep3}^{(+)}$&$\scr{\ep1\ep3\ep1\ep3}$&$\scr{\ep1\ep3\ep1\ep1}^{(-)}$&$\scr{\ep1\ep1}$\\
\hline
$\scr{\ep3\ep1}$&$\scr{\ep3\ep1\ep3\ep1}$&$\scr{\ep3\ep1\ep3\ep3}^{(+)}$&$\scr{\ep3\ep3\ep3\ep3}$&$\scr{\ep3\ep3\ep1\ep3}^{(+)}$&$\scr{\ep1\ep3\ep1\ep3}$&$\scr{\ep1\ep3\ep1\ep1}^{(-)}$&$\scr{\ep1\ep1}$\\
\hline
$\scr{\ep4\ep3}$&$\scr{\ep4\ep3\ep4\ep3}$&$\scr{\ep4\ep3\ep3\ep3}^{(+)}$&$\scr{\ep3\ep3\ep3\ep3}$&$\scr{\ep3\ep3\ep1\ep3}^{(+)}$&$\scr{\ep1\ep3\ep1\ep3}$&$\scr{\ep1\ep3\ep1\ep1}^{(-)}$&$\scr{\ep1\ep1}$\\
\hline
$\scr{\ep5\ep3}$&$\scr{\ep5\ep3\ep5\ep3}$&$\scr{\ep5\ep3\ep3\ep3}^{(+)}$&$\scr{\ep3\ep3\ep3\ep3}$&$\scr{\ep3\ep3\ep1\ep3}^{(+)}$&$\scr{\ep1\ep3\ep1\ep3}$&$\scr{\ep1\ep3\ep1\ep1}^{(-)}$&$\scr{\ep1\ep1}$\\
\hline
$\scr{\ep6\ep3}$&$\scr{\ep6\ep3\ep6\ep3}$&$\scr{\ep6\ep3\ep3\ep3}^{(+)}$&$\scr{\ep3\ep3\ep3\ep3}$&$\scr{\ep3\ep3\ep1\ep3}^{(+)}$&$\scr{\ep1\ep3\ep1\ep3}$&$\scr{\ep1\ep3\ep1\ep1}^{(-)}$&$\scr{\ep1\ep1}$\\
\hline
$\scr{\ep2\ep1}$&$\scr{\ep2\ep1\ep2\ep1}$&$\scr{\ep2\ep1\ep2\ep3}^{(+)}$&$\scr{\ep2\ep3\ep2\ep3}$&$\scr{\ep2\ep3\ep1\ep3}^{(+)}$&$\scr{\ep1\ep3\ep1\ep3}$&$\scr{\ep1\ep3\ep1\ep1}^{(-)}$&$\scr{\ep1\ep1}$\\
\hline
$\scr{\ep4\ep3}$&$\scr{\ep4\ep3\ep4\ep3}$&$\scr{\ep4\ep3\ep2\ep3}^{(+)}$&$\scr{\ep2\ep3\ep2\ep3}$&$\scr{\ep2\ep3\ep1\ep3}^{(+)}$&$\scr{\ep1\ep3\ep1\ep3}$&$\scr{\ep1\ep3\ep1\ep1}^{(-)}$&$\scr{\ep1\ep1}$\\
\hline
$\scr{\ep5\ep3}$&$\scr{\ep5\ep3\ep5\ep3}$&$\scr{\ep5\ep3\ep2\ep3}^{(+)}$&$\scr{\ep2\ep3\ep2\ep3}$&$\scr{\ep2\ep3\ep1\ep3}^{(+)}$&$\scr{\ep1\ep3\ep1\ep3}$&$\scr{\ep1\ep3\ep1\ep1}^{(-)}$&$\scr{\ep1\ep1}$\\
\hline
$\scr{\ep6\ep3}$&$\scr{\ep6\ep3\ep6\ep3}$&$\scr{\ep6\ep3\ep2\ep3}^{(+)}$&$\scr{\ep2\ep3\ep2\ep3}$&$\scr{\ep2\ep3\ep1\ep3}^{(+)}$&$\scr{\ep1\ep3\ep1\ep3}$&$\scr{\ep1\ep3\ep1\ep1}^{(-)}$&$\scr{\ep1\ep1}$\\
\hline
$\scr{\ep2\ep1}$&$\scr{\ep2\ep1\ep2\ep1}$&$\scr{\ep2\ep1\ep1\ep1}^{(+)}$&$\scr{\ep1\ep1\ep1\ep1}$&$\scr{\ep1\ep1\ep2\ep1}^{(-)}$&$\scr{\ep2\ep1\ep2\ep1}$&$\scr{\ep2\ep1\ep1\ep1}^{(+)}$&$\scr{\ep1\ep1}$\\
\hline
$\scr{\ep3\ep1}$&$\scr{\ep3\ep1\ep3\ep1}$&$\scr{\ep3\ep1\ep1\ep1}^{(+)}$&$\scr{\ep1\ep1\ep1\ep1}$&$\scr{\ep1\ep1\ep2\ep1}^{(-)}$&$\scr{\ep2\ep1\ep2\ep1}$&$\scr{\ep2\ep1\ep1\ep1}^{(+)}$&$\scr{\ep1\ep1}$\\
\hline
$\scr{\ep2\ep1}$&$\scr{\ep2\ep1\ep2\ep1}$&$\scr{\ep2\ep1\ep2\ep2}^{(+)}$&$\scr{\ep2\ep2\ep2\ep2}$&$\scr{\ep2\ep2\ep2\ep1}^{(-)}$&$\scr{\ep2\ep1\ep2\ep1}$&$\scr{\ep2\ep1\ep1\ep1}^{(+)}$&$\scr{\ep1\ep1}$\\
\hline
$\scr{\ep4\ep2}$&$\scr{\ep4\ep2\ep4\ep2}$&$\scr{\ep4\ep2\ep2\ep2}^{(+)}$&$\scr{\ep2\ep2\ep2\ep2}$&$\scr{\ep2\ep2\ep2\ep1}^{(-)}$&$\scr{\ep2\ep1\ep2\ep1}$&$\scr{\ep2\ep1\ep1\ep1}^{(+)}$&$\scr{\ep1\ep1}$\\
\hline
$\scr{\ep5\ep2}$&$\scr{\ep5\ep2\ep5\ep2}$&$\scr{\ep5\ep2\ep2\ep2}^{(+)}$&$\scr{\ep2\ep2\ep2\ep2}$&$\scr{\ep2\ep2\ep2\ep1}^{(-)}$&$\scr{\ep2\ep1\ep2\ep1}$&$\scr{\ep2\ep1\ep1\ep1}^{(+)}$&$\scr{\ep1\ep1}$\\
\hline
$\scr{\ep6\ep2}$&$\scr{\ep6\ep2\ep6\ep2}$&$\scr{\ep6\ep2\ep2\ep2}^{(+)}$&$\scr{\ep2\ep2\ep2\ep2}$&$\scr{\ep2\ep2\ep2\ep1}^{(-)}$&$\scr{\ep2\ep1\ep2\ep1}$&$\scr{\ep2\ep1\ep1\ep1}^{(+)}$&$\scr{\ep1\ep1}$\\
\hline
$\scr{\ep2\ep1}$&$\scr{\ep2\ep1\ep2\ep1}$&$\scr{\ep2\ep1\ep2\ep3}^{(+)}$&$\scr{\ep2\ep3\ep2\ep3}$&$\scr{\ep2\ep3\ep2\ep1}^{(-)}$&$\scr{\ep2\ep1\ep2\ep1}$&$\scr{\ep2\ep1\ep1\ep1}^{(+)}$&$\scr{\ep1\ep1}$\\
\hline
$\scr{\ep4\ep3}$&$\scr{\ep4\ep3\ep4\ep3}$&$\scr{\ep4\ep3\ep2\ep3}^{(+)}$&$\scr{\ep2\ep3\ep2\ep3}$&$\scr{\ep2\ep3\ep2\ep1}^{(-)}$&$\scr{\ep2\ep1\ep2\ep1}$&$\scr{\ep2\ep1\ep1\ep1}^{(+)}$&$\scr{\ep1\ep1}$\\
\hline
$\scr{\ep5\ep3}$&$\scr{\ep5\ep3\ep5\ep3}$&$\scr{\ep5\ep3\ep2\ep3}^{(+)}$&$\scr{\ep2\ep3\ep2\ep3}$&$\scr{\ep2\ep3\ep2\ep1}^{(-)}$&$\scr{\ep2\ep1\ep2\ep1}$&$\scr{\ep2\ep1\ep1\ep1}^{(+)}$&$\scr{\ep1\ep1}$\\
\hline
$\scr{\ep6\ep3}$&$\scr{\ep6\ep3\ep6\ep3}$&$\scr{\ep6\ep3\ep2\ep3}^{(+)}$&$\scr{\ep2\ep3\ep2\ep3}$&$\scr{\ep2\ep3\ep2\ep1}^{(-)}$&$\scr{\ep2\ep1\ep2\ep1}$&$\scr{\ep2\ep1\ep1\ep1}^{(+)}$&$\scr{\ep1\ep1}$\\
\hline
$\scr{\ep4\ep2}$&$\scr{\ep4\ep2\ep4\ep2}$&$\scr{\ep4\ep2\ep4\ep1}^{(-)}$&$\scr{\ep4\ep1\ep4\ep1}$&$\scr{\ep4\ep1\ep2\ep1}^{(+)}$&$\scr{\ep2\ep1\ep2\ep1}$&$\scr{\ep2\ep1\ep1\ep1}^{(+)}$&$\scr{\ep1\ep1}$\\
\hline
$\scr{\ep4\ep3}$&$\scr{\ep4\ep3\ep4\ep3}$&$\scr{\ep4\ep3\ep4\ep1}^{(-)}$&$\scr{\ep4\ep1\ep4\ep1}$&$\scr{\ep4\ep1\ep2\ep1}^{(+)}$&$\scr{\ep2\ep1\ep2\ep1}$&$\scr{\ep2\ep1\ep1\ep1}^{(+)}$&$\scr{\ep1\ep1}$\\
\hline
$\scr{\ep2\ep1}$&$\scr{\ep2\ep1\ep2\ep1}$&$\scr{\ep2\ep1\ep4\ep1}^{(-)}$&$\scr{\ep4\ep1\ep4\ep1}$&$\scr{\ep4\ep1\ep2\ep1}^{(+)}$&$\scr{\ep2\ep1\ep2\ep1}$&$\scr{\ep2\ep1\ep1\ep1}^{(+)}$&$\scr{\ep1\ep1}$\\
\hline
$\scr{\ep3\ep1}$&$\scr{\ep3\ep1\ep3\ep1}$&$\scr{\ep3\ep1\ep4\ep1}^{(-)}$&$\scr{\ep4\ep1\ep4\ep1}$&$\scr{\ep4\ep1\ep2\ep1}^{(+)}$&$\scr{\ep2\ep1\ep2\ep1}$&$\scr{\ep2\ep1\ep1\ep1}^{(+)}$&$\scr{\ep1\ep1}$\\
\hline
$\scr{\ep5\ep2}$&$\scr{\ep5\ep2\ep5\ep2}$&$\scr{\ep5\ep2\ep5\ep1}^{(-)}$&$\scr{\ep5\ep1\ep5\ep1}$&$\scr{\ep5\ep1\ep2\ep1}^{(+)}$&$\scr{\ep2\ep1\ep2\ep1}$&$\scr{\ep2\ep1\ep1\ep1}^{(+)}$&$\scr{\ep1\ep1}$\\
\hline
$\scr{\ep5\ep3}$&$\scr{\ep5\ep3\ep5\ep3}$&$\scr{\ep5\ep3\ep5\ep1}^{(-)}$&$\scr{\ep5\ep1\ep5\ep1}$&$\scr{\ep5\ep1\ep2\ep1}^{(+)}$&$\scr{\ep2\ep1\ep2\ep1}$&$\scr{\ep2\ep1\ep1\ep1}^{(+)}$&$\scr{\ep1\ep1}$\\
\hline
$\scr{\ep2\ep1}$&$\scr{\ep2\ep1\ep2\ep1}$&$\scr{\ep2\ep1\ep5\ep1}^{(-)}$&$\scr{\ep5\ep1\ep5\ep1}$&$\scr{\ep5\ep1\ep2\ep1}^{(+)}$&$\scr{\ep2\ep1\ep2\ep1}$&$\scr{\ep2\ep1\ep1\ep1}^{(+)}$&$\scr{\ep1\ep1}$\\
\hline
$\scr{\ep3\ep1}$&$\scr{\ep3\ep1\ep3\ep1}$&$\scr{\ep3\ep1\ep5\ep1}^{(-)}$&$\scr{\ep5\ep1\ep5\ep1}$&$\scr{\ep5\ep1\ep2\ep1}^{(+)}$&$\scr{\ep2\ep1\ep2\ep1}$&$\scr{\ep2\ep1\ep1\ep1}^{(+)}$&$\scr{\ep1\ep1}$\\
\hline
$\scr{\ep6\ep2}$&$\scr{\ep6\ep2\ep6\ep2}$&$\scr{\ep6\ep2\ep6\ep1}^{(-)}$&$\scr{\ep6\ep1\ep6\ep1}$&$\scr{\ep6\ep1\ep2\ep1}^{(+)}$&$\scr{\ep2\ep1\ep2\ep1}$&$\scr{\ep2\ep1\ep1\ep1}^{(+)}$&$\scr{\ep1\ep1}$\\
\hline
$\scr{\ep6\ep3}$&$\scr{\ep6\ep3\ep6\ep3}$&$\scr{\ep6\ep3\ep6\ep1}^{(-)}$&$\scr{\ep6\ep1\ep6\ep1}$&$\scr{\ep6\ep1\ep2\ep1}^{(+)}$&$\scr{\ep2\ep1\ep2\ep1}$&$\scr{\ep2\ep1\ep1\ep1}^{(+)}$&$\scr{\ep1\ep1}$\\
\hline
$\scr{\ep2\ep1}$&$\scr{\ep2\ep1\ep2\ep1}$&$\scr{\ep2\ep1\ep6\ep1}^{(-)}$&$\scr{\ep6\ep1\ep6\ep1}$&$\scr{\ep6\ep1\ep2\ep1}^{(+)}$&$\scr{\ep2\ep1\ep2\ep1}$&$\scr{\ep2\ep1\ep1\ep1}^{(+)}$&$\scr{\ep1\ep1}$\\
\hline
$\scr{\ep3\ep1}$&$\scr{\ep3\ep1\ep3\ep1}$&$\scr{\ep3\ep1\ep6\ep1}^{(-)}$&$\scr{\ep6\ep1\ep6\ep1}$&$\scr{\ep6\ep1\ep2\ep1}^{(+)}$&$\scr{\ep2\ep1\ep2\ep1}$&$\scr{\ep2\ep1\ep1\ep1}^{(+)}$&$\scr{\ep1\ep1}$\\
\hline
$\scr{\ep2\ep1}$&$\scr{\ep2\ep1\ep2\ep1}$&$\scr{\ep2\ep1\ep1\ep1}^{(+)}$&$\scr{\ep1\ep1\ep1\ep1}$&$\scr{\ep1\ep1\ep3\ep1}^{(-)}$&$\scr{\ep3\ep1\ep3\ep1}$&$\scr{\ep3\ep1\ep1\ep1}^{(+)}$&$\scr{\ep1\ep1}$\\
\hline
$\scr{\ep3\ep1}$&$\scr{\ep3\ep1\ep3\ep1}$&$\scr{\ep3\ep1\ep1\ep1}^{(+)}$&$\scr{\ep1\ep1\ep1\ep1}$&$\scr{\ep1\ep1\ep3\ep1}^{(-)}$&$\scr{\ep3\ep1\ep3\ep1}$&$\scr{\ep3\ep1\ep1\ep1}^{(+)}$&$\scr{\ep1\ep1}$\\
\hline
$\scr{\ep3\ep1}$&$\scr{\ep3\ep1\ep3\ep1}$&$\scr{\ep3\ep1\ep3\ep2}^{(+)}$&$\scr{\ep3\ep2\ep3\ep2}$&$\scr{\ep3\ep2\ep3\ep1}^{(-)}$&$\scr{\ep3\ep1\ep3\ep1}$&$\scr{\ep3\ep1\ep1\ep1}^{(+)}$&$\scr{\ep1\ep1}$\\
\hline
$\scr{\ep4\ep2}$&$\scr{\ep4\ep2\ep4\ep2}$&$\scr{\ep4\ep2\ep3\ep2}^{(+)}$&$\scr{\ep3\ep2\ep3\ep2}$&$\scr{\ep3\ep2\ep3\ep1}^{(-)}$&$\scr{\ep3\ep1\ep3\ep1}$&$\scr{\ep3\ep1\ep1\ep1}^{(+)}$&$\scr{\ep1\ep1}$\\
\hline
$\scr{\ep5\ep2}$&$\scr{\ep5\ep2\ep5\ep2}$&$\scr{\ep5\ep2\ep3\ep2}^{(+)}$&$\scr{\ep3\ep2\ep3\ep2}$&$\scr{\ep3\ep2\ep3\ep1}^{(-)}$&$\scr{\ep3\ep1\ep3\ep1}$&$\scr{\ep3\ep1\ep1\ep1}^{(+)}$&$\scr{\ep1\ep1}$\\
\hline
$\scr{\ep6\ep2}$&$\scr{\ep6\ep2\ep6\ep2}$&$\scr{\ep6\ep2\ep3\ep2}^{(+)}$&$\scr{\ep3\ep2\ep3\ep2}$&$\scr{\ep3\ep2\ep3\ep1}^{(-)}$&$\scr{\ep3\ep1\ep3\ep1}$&$\scr{\ep3\ep1\ep1\ep1}^{(+)}$&$\scr{\ep1\ep1}$\\
\hline
$\scr{\ep3\ep1}$&$\scr{\ep3\ep1\ep3\ep1}$&$\scr{\ep3\ep1\ep3\ep3}^{(+)}$&$\scr{\ep3\ep3\ep3\ep3}$&$\scr{\ep3\ep3\ep3\ep1}^{(-)}$&$\scr{\ep3\ep1\ep3\ep1}$&$\scr{\ep3\ep1\ep1\ep1}^{(+)}$&$\scr{\ep1\ep1}$\\
\hline
$\scr{\ep4\ep3}$&$\scr{\ep4\ep3\ep4\ep3}$&$\scr{\ep4\ep3\ep3\ep3}^{(+)}$&$\scr{\ep3\ep3\ep3\ep3}$&$\scr{\ep3\ep3\ep3\ep1}^{(-)}$&$\scr{\ep3\ep1\ep3\ep1}$&$\scr{\ep3\ep1\ep1\ep1}^{(+)}$&$\scr{\ep1\ep1}$\\
\hline
$\scr{\ep5\ep3}$&$\scr{\ep5\ep3\ep5\ep3}$&$\scr{\ep5\ep3\ep3\ep3}^{(+)}$&$\scr{\ep3\ep3\ep3\ep3}$&$\scr{\ep3\ep3\ep3\ep1}^{(-)}$&$\scr{\ep3\ep1\ep3\ep1}$&$\scr{\ep3\ep1\ep1\ep1}^{(+)}$&$\scr{\ep1\ep1}$\\
\hline
$\scr{\ep6\ep3}$&$\scr{\ep6\ep3\ep6\ep3}$&$\scr{\ep6\ep3\ep3\ep3}^{(+)}$&$\scr{\ep3\ep3\ep3\ep3}$&$\scr{\ep3\ep3\ep3\ep1}^{(-)}$&$\scr{\ep3\ep1\ep3\ep1}$&$\scr{\ep3\ep1\ep1\ep1}^{(+)}$&$\scr{\ep1\ep1}$\\
\hline
$\scr{\ep4\ep2}$&$\scr{\ep4\ep2\ep4\ep2}$&$\scr{\ep4\ep2\ep4\ep1}^{(-)}$&$\scr{\ep4\ep1\ep4\ep1}$&$\scr{\ep4\ep1\ep3\ep1}^{(+)}$&$\scr{\ep3\ep1\ep3\ep1}$&$\scr{\ep3\ep1\ep1\ep1}^{(+)}$&$\scr{\ep1\ep1}$\\
\hline
$\scr{\ep4\ep3}$&$\scr{\ep4\ep3\ep4\ep3}$&$\scr{\ep4\ep3\ep4\ep1}^{(-)}$&$\scr{\ep4\ep1\ep4\ep1}$&$\scr{\ep4\ep1\ep3\ep1}^{(+)}$&$\scr{\ep3\ep1\ep3\ep1}$&$\scr{\ep3\ep1\ep1\ep1}^{(+)}$&$\scr{\ep1\ep1}$\\
\hline
$\scr{\ep2\ep1}$&$\scr{\ep2\ep1\ep2\ep1}$&$\scr{\ep2\ep1\ep4\ep1}^{(-)}$&$\scr{\ep4\ep1\ep4\ep1}$&$\scr{\ep4\ep1\ep3\ep1}^{(+)}$&$\scr{\ep3\ep1\ep3\ep1}$&$\scr{\ep3\ep1\ep1\ep1}^{(+)}$&$\scr{\ep1\ep1}$\\
\hline
$\scr{\ep3\ep1}$&$\scr{\ep3\ep1\ep3\ep1}$&$\scr{\ep3\ep1\ep4\ep1}^{(-)}$&$\scr{\ep4\ep1\ep4\ep1}$&$\scr{\ep4\ep1\ep3\ep1}^{(+)}$&$\scr{\ep3\ep1\ep3\ep1}$&$\scr{\ep3\ep1\ep1\ep1}^{(+)}$&$\scr{\ep1\ep1}$\\
\hline
$\scr{\ep5\ep2}$&$\scr{\ep5\ep2\ep5\ep2}$&$\scr{\ep5\ep2\ep5\ep1}^{(-)}$&$\scr{\ep5\ep1\ep5\ep1}$&$\scr{\ep5\ep1\ep3\ep1}^{(+)}$&$\scr{\ep3\ep1\ep3\ep1}$&$\scr{\ep3\ep1\ep1\ep1}^{(+)}$&$\scr{\ep1\ep1}$\\
\hline
$\scr{\ep5\ep3}$&$\scr{\ep5\ep3\ep5\ep3}$&$\scr{\ep5\ep3\ep5\ep1}^{(-)}$&$\scr{\ep5\ep1\ep5\ep1}$&$\scr{\ep5\ep1\ep3\ep1}^{(+)}$&$\scr{\ep3\ep1\ep3\ep1}$&$\scr{\ep3\ep1\ep1\ep1}^{(+)}$&$\scr{\ep1\ep1}$\\
\hline
$\scr{\ep2\ep1}$&$\scr{\ep2\ep1\ep2\ep1}$&$\scr{\ep2\ep1\ep5\ep1}^{(-)}$&$\scr{\ep5\ep1\ep5\ep1}$&$\scr{\ep5\ep1\ep3\ep1}^{(+)}$&$\scr{\ep3\ep1\ep3\ep1}$&$\scr{\ep3\ep1\ep1\ep1}^{(+)}$&$\scr{\ep1\ep1}$\\
\hline
$\scr{\ep3\ep1}$&$\scr{\ep3\ep1\ep3\ep1}$&$\scr{\ep3\ep1\ep5\ep1}^{(-)}$&$\scr{\ep5\ep1\ep5\ep1}$&$\scr{\ep5\ep1\ep3\ep1}^{(+)}$&$\scr{\ep3\ep1\ep3\ep1}$&$\scr{\ep3\ep1\ep1\ep1}^{(+)}$&$\scr{\ep1\ep1}$\\
\hline
$\scr{\ep6\ep2}$&$\scr{\ep6\ep2\ep6\ep2}$&$\scr{\ep6\ep2\ep6\ep1}^{(-)}$&$\scr{\ep6\ep1\ep6\ep1}$&$\scr{\ep6\ep1\ep3\ep1}^{(+)}$&$\scr{\ep3\ep1\ep3\ep1}$&$\scr{\ep3\ep1\ep1\ep1}^{(+)}$&$\scr{\ep1\ep1}$\\
\hline
$\scr{\ep6\ep3}$&$\scr{\ep6\ep3\ep6\ep3}$&$\scr{\ep6\ep3\ep6\ep1}^{(-)}$&$\scr{\ep6\ep1\ep6\ep1}$&$\scr{\ep6\ep1\ep3\ep1}^{(+)}$&$\scr{\ep3\ep1\ep3\ep1}$&$\scr{\ep3\ep1\ep1\ep1}^{(+)}$&$\scr{\ep1\ep1}$\\
\hline
$\scr{\ep2\ep1}$&$\scr{\ep2\ep1\ep2\ep1}$&$\scr{\ep2\ep1\ep6\ep1}^{(-)}$&$\scr{\ep6\ep1\ep6\ep1}$&$\scr{\ep6\ep1\ep3\ep1}^{(+)}$&$\scr{\ep3\ep1\ep3\ep1}$&$\scr{\ep3\ep1\ep1\ep1}^{(+)}$&$\scr{\ep1\ep1}$\\
\hline
$\scr{\ep3\ep1}$&$\scr{\ep3\ep1\ep3\ep1}$&$\scr{\ep3\ep1\ep6\ep1}^{(-)}$&$\scr{\ep6\ep1\ep6\ep1}$&$\scr{\ep6\ep1\ep3\ep1}^{(+)}$&$\scr{\ep3\ep1\ep3\ep1}$&$\scr{\ep3\ep1\ep1\ep1}^{(+)}$&$\scr{\ep1\ep1}$\\
\hline
\hline
\end{supertabular}}
\end{center}

\include{figures}
\include{captions}
\newpage

\end{document}